\documentclass[preprint2]{aastex}

\usepackage[usenames]{color}

 \def\gtsima{$\; \buildrel > \over \sim \;$}
 \def\simgt{\lower.5ex\hbox{\gtsima}}

\shorttitle{Suzaku observations of subhalos in the coma cluster}
\shortauthors{Sasaki et al.}

\begin{document}


\title{SUZAKU OBSERVATIONS OF SUBHALOS IN THE COMA CLUSTER}


\author{
Toru \textsc{Sasaki}\altaffilmark{1}, 
Kyoko \textsc{Matsushita}\altaffilmark{1}, 
Kosuke \textsc{Sato}\altaffilmark{1}, 
and Nobuhiro \textsc{Okabe}\altaffilmark{2,3}
}



\altaffiltext{1}{Department of Physics, Tokyo University of Science, 
1-3 Kagurazaka, Shinjuku-ku, Tokyo 162-8601, Japan}
\altaffiltext{2}{Department of Physical Science, Hiroshima University, 1-3-1 Kagamiyama, Higashi-Hiroshima, Hiroshima 739-8526, Japan}
\altaffiltext{3}{Kavli Institute for the Physics and Mathematics of the Universe (WPI), \\Todai Institutes for Advanced Study,University of Tokyo, 5-1-5 Kashiwanoha, Kashiwa, Chiba 277-8583, Japan}
\email{j1213703@ed.tus.ac.jp; matusita@rs.kagu.tus.ac.jp}


\begin{abstract}

We observed three massive subhalos in the Coma cluster with {\it Suzaku}.
These subhalos, labeled ``ID~1", ``ID~2", and ``ID~32",
  were detected with a weak-lensing survey using the Subaru/Suprime-Cam \citep{Okabe2013}, and
are located at 
the projected distances of 1.4~$r_{500}$,  1.2~$r_{500}$, and  1.6~$r_{500}$ 
from the center of the Coma cluster, respectively.
The subhalo ``ID~1" has a compact X-ray excess emission close to the center of 
the weak-lensing mass contour, and the gas mass to weak-lensing mass ratio is about 0.001.
The temperature of the emission is about 3~keV,
which is slightly lower than that of the surrounding intracluster medium (ICM) and 
that expected for  the temperature vs. mass relation of clusters of galaxies.
The subhalo ``ID~32" shows an excess emission whose peak 
is shifted toward the opposite direction from the center of the Coma cluster. 
The gas mass to weak-lensing mass ratio is also about 0.001, 
which is significantly smaller than regular galaxy groups.
The temperature of the excess is about 0.5~keV and  significantly lower than 
that of the surrounding ICM and far from the temperature vs. mass relation of clusters.
However, there is no significant excess X-ray  emission in the ``ID~2" subhalo. 
Assuming an infall velocity of about 2000 $\rm km~s^{-1}$, at the border of
the excess X-ray emission, the ram pressures for ``ID~1" and ``ID~32" are comparable
to the gravitational restoring force per area.
We also studied  the effect of the Kelvin-Helmholtz instability to strip the gas.
Although we found X-ray clumps associated with the weak-lensing subhalos,
their X-ray luminosities are much lower than the total ICM luminosity
in the cluster outskirts.
\end{abstract}


\section{INTRODUCTION}
Galaxy clusters are the largest self-gravitating bound systems
in the Universe and composed of thousands of galaxies, 
the intracluster medium (ICM), and the dark matter.
The ICM covers the total mass range of roughly $10^{13-14}~M_{\sun}$ 
and is bound to the potential of the dark matter halo 
which covers that of $10^{14-15}~M_{\sun}$.
Therefore, the gravity of the dark matter halo plays the most 
important role in cluster evolution and structure formation. 
Numerical simulations with the Cold Dark Matter (CDM) model 
predict that the galaxy clusters form through merger or accretion of 
the smaller system like galaxy groups.
Since the dynamical time scale of galaxy clusters is comparable 
to the Hubble time, outskirts of the galaxy clusters still maintain 
the evolution effects via the accretion of the substructures.
The central region of these accreting objects is expected to be survived until recent days
as subhalos in the cluster host halo.

The mass distribution of subhalos provides us with information on 
 the mass assembly of galaxy cluster.
\citet{Okabe2013} surveyed and measured the mass of subhalos 
in the Coma cluster using  weak-lensing observations  with the Subaru/Suprime-Cam.
Thanks to the large apparent size, they detected 32 cluster subhalos 
whose mass range is $\sim 2-50\times10^{12}~h^{-1}M_{\sun}$.
They first confirmed that the subhalo mass function, 
$d~n/d~ln~M_{\rm sub}$, is well represented with 
a single powerlaw or a Schechter function. 
The best-fit indexes of each model are $\sim1$, which 
agree well with CDM model prediction on sub-scale of cluster.
Stacked signals of subhalos were well represented with a sharply truncated 
Navarro-Frenk-White (NFW) mass model \citep{Navarro1995} as expected from a tidal destruction model.
For three most massive subhalos whose mass are higher than 
$\sim1\times10^{13}~h^{-1}M_{\sun}$, 
they measured mass and truncation radius of each subhalo.
One of them is a famous substructure  of the Coma cluster around the NGC~4839 group.


If subhalos in the cluster outskirts still possess some amount of their hot gas
and are not excluded in X-ray analysis, the derived ICM density would be overestimated.
The recent $Suzaku$ observations reported that 
the entropy of the ICM, which is a useful parameter for the
thermodynamical history,  shows flatter profiles beyond $r_{500}$ 
 than expectations from pure gravitational heating
\citep{George2009, Reiprich2009, Bautz2009, Hoshino2010, Kawaharada2010, Simionescu2011, Urban2011, Akamatsu2011, Walker2012a, Akamatsu2012, Walker2012b, SatoT2012, Walker2013, Ichikawa2013, Simionescu2013, Urban2014, Sato2014, Okabe2014}.
Since the gas fraction in the Perseus cluster assumed hydrostatic equilibrium 
exceeds the cosmic baryon fraction, \citet{Simionescu2011} claimed the 
effect of the gas clumpiness, although there is no significant excess 
X-ray sources in the outskirts with a Chandra observation \citep{Urban2014}.

Weak-lensing mass estimation of cluster main halos are complementary to X-ray observations.
In the cluster outskirts,
the hydrostatic mass with {\it Suzaku}  is significantly lower
than the weak-lensing mass with the Subaru telescope 
\citep{Kawaharada2010, Ichikawa2013,Mochizuki2014, Okabe2014}.
\citet{Okabe2014} discussed that
the bivariate scaling functions of the electron density and temperature 
indicate that entropy flattening of the outskirts of the galaxy clusters 
caused by the steepening of  temperature profiles.
Deviations from hydrostatic equilibrium have been discussed in
\citet{Kawaharada2010, Ichikawa2013, Mochizuki2014, Okabe2014}.
\citet{Hoshino2010} proposed an another idea 
that electron temperature is lower than the ion temperature in these regions,
since heating the electrons takes a longer time than that of the ion 
after accretion shocks and mergers.   

With weak-lensing mass measurements of subhalos, 
 we can  search X-ray clumps associated with these subhalos efficiently with X-ray observations.
In this paper, we describe the X-ray properties of three massive subhalos, 
whose mass is greater than $\sim 9\times 10^{12}~h^{-1}M_{\sun}$, 
detected Subaru weak-lensing observations of the Coma cluster \citep{Okabe2013}.
Excluding the NGC~4839 subgroup, we observed two of the most massive subhalos with $Suzaku$.
We also observed a smaller subhalo, whose total mass is $\sim 9\times 10^{12}~h^{-1}M_{\sun}$.
We summarize the observations and date preparation in section \ref{sec:obs}.
Section \ref{sec:surface} shows the X-ray images and surface profiles of each subhalo.
In section \ref{sec:results}, we summarize the spectral fitting and that results.
We compare of the X-ray properties with those of other galaxy groups,
and discuss the effect of ram pressure and gas clumpiness in section \ref{sec:discussion}.

We use $\Omega_{m,0}=0.27, \Omega_{\Lambda}=0.73$, 
and $H_{0}=$70~km~s$^{-1}$~Mpc$^{-1}$ in this paper.
At the redshift of the cluster, $z$ = 0.0231 \citep{Struble1999}, 
1~arcmin corresponds to 28.9~kpc.
The solar abundance table is given by \citet{Lodders2003}. The 
errors are in the 68\% confidence region for the single parameter 
of interest.

\section{OBSERVATION AND DATA REDUCTION}
\label{sec:obs}

\begin{deluxetable}{lllllll}
\tabletypesize{\scriptsize}
\tablewidth{0pt}
\tablecaption{
Properties of the Coma cluster subhalos.
\label{tb:subhalodata}
}
\tablehead{
ID$^{a}$ &  $M_{2D}^{b}$ &  $M^{c}$ & $r_{t}^{d}$ &  (R.A., decl.)$^{e}$   & $N_{H}^{f}$ & \colhead{Distance$^{g}$}  \\
& $10^{12}~h^{-1}~M_{\sun}$ & $10^{12}~h^{-1}~M_{\sun}$ & arcmin  &  J2000.0 & $10^{19}~{\rm cm}^{-2}$ & \colhead{Arcmin/$r_{500}$} }
\startdata
1  & $15.42\pm2.79$ & $14.26_{-2.53-5.55}^{+2.37}$ & ${3.86}_{-0.19}^{+0.14}$ &  $12^{\rm h}55^{\rm m}34\fs5, +27^{\circ}31\arcmin33.7\arcsec$  & 8.6 & 61.8/1.42\\
2 &  $8.79\pm4.69$ & - & - & $12^{\rm h}56^{\rm m}03\fs8, +27^{\circ}47\arcmin20.8\arcsec$  & 8.7 & 51.6/1.18\\
32 & $45.95\pm7.57$ & $47.75_{-5.81-13.42}^{+5.81}$    & ${9.21}_{-0.83}^{+0.74}$ & $13^{\rm h}01^{\rm m}41\fs0, +29^{\circ}03\arcmin14.4\arcsec$  & 9.5 & 71.2/1.63\\
\enddata
\tablenotetext{a}{The name of subhalos \citep{Okabe2013}.}
\tablenotetext{b}{The projected weak lens mass of the subhalos \citep{Okabe2013}. }
\tablenotetext{c}{The best-fit mass with truncated NFW model \citep{Okabe2013}. }
\tablenotetext{d}{The truncation radius derived from \citep{Okabe2013}.}
\tablenotetext{e}{
For ``ID~1" and ``ID~2", the center of the subhalos determined from the mass contour. 
The center of ``ID32", however, is derived from the weak-lensing signal peak since 
the difference between weak-lensing signal peak and mass contour is 
significantly smaller than the {\it Suzaku} point spread function.
}
\tablenotetext{f}{The Galactic hydrogen column density \citep{Kalberla2005}.}
\tablenotetext{g}{The distance from the X-ray peak of the Coma cluster center ($12^{\rm h}59^{\rm m}44\fs81, 27{\degr}56{\arcmin}49\farcs92$).}
\end{deluxetable}

\begin{deluxetable}{lllll}
\tabletypesize{\scriptsize}
\tablewidth{0pt}
\tablecaption{
$Suzaku$ observation logs for Coma cluster subhalos.
\label{tb:obslog}
}
\tablehead{
\colhead{Field name} & \colhead{Sequence} &  \colhead{Date-Obs.$^{a}$} & \colhead{(R.A., decl.)$^{b}$}  &  \colhead{Exposure$^{c}$} \\
\colhead{} & \colhead{Number} & \colhead{} & \colhead{J2000.0} & \colhead{ksec} }
\startdata
ID~1             & 808022010 & 2013-06-10T14:12:00 & $12^{\rm h}55^{\rm m}28\fs0$, $27\degr31\arcmin00\farcs1$ & 18.4 \\  
ID~2             & 808021010 & 2013-06-10T00:39:15 &$12^{\rm h}55^{\rm m}55\fs3$, $27\degr45\arcmin17\farcs6$ & 23.7 \\
ID~32           & 808018010 & 2013-06-08T09:04:42 &$13^{\rm h}01^{\rm m}36\fs1$, $29\degr01\arcmin40\farcs8$ & 26.4 \\
ID~32 BGD  & 808019010 & 2013-06-09T01:38:51 & $13^{\rm h}01^{\rm m}00\fs7$, $28\degr45\arcmin46\farcs4$ & 20.2 \\ 
\enddata
\tablenotetext{a}{Start date of observation, written in the DATE-OBS keyword of the event FITS files.}
\tablenotetext{b}{Average pointing direction of the XIS, written in the RA\_NOM and DEC\_NOM keywords of the event FITS files.}
\tablenotetext{c}{Exposure time after screening.}
\end{deluxetable}

\begin{figure*}[htpd]
\begin{center}
\includegraphics[width=0.80\textwidth,angle=90,clip]{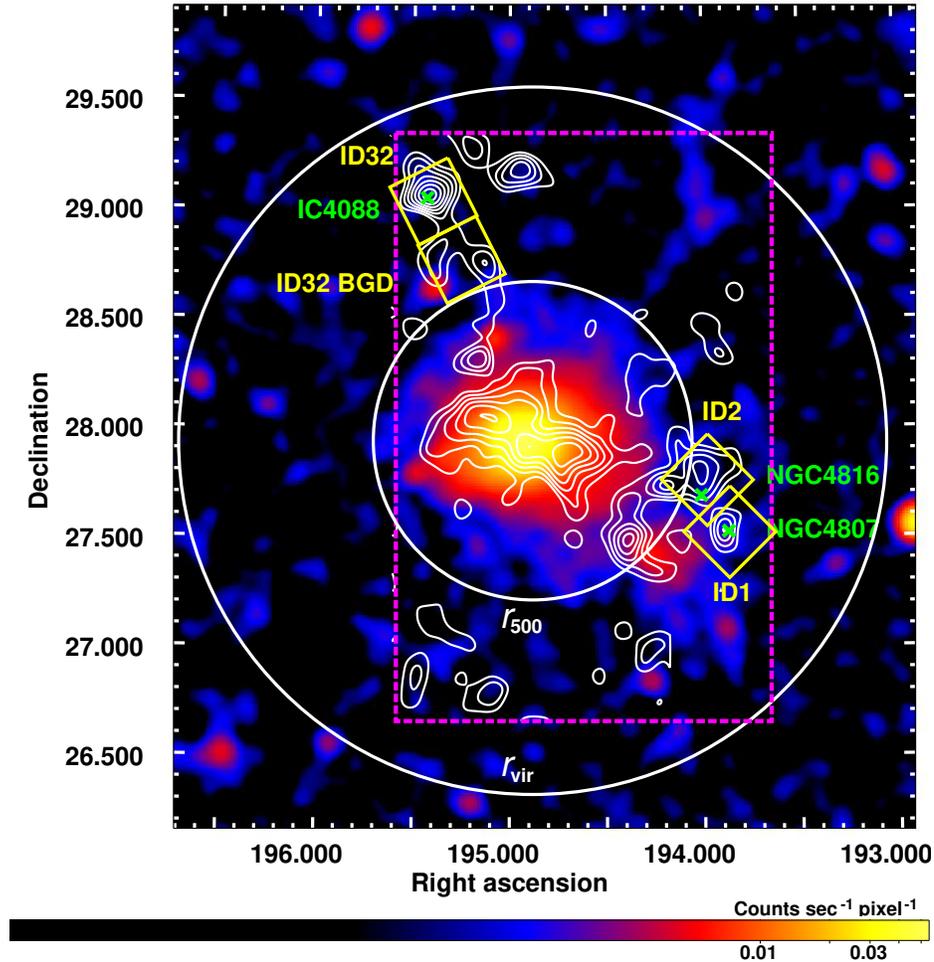}
\caption{
The weak-lensing mass contour map in linear scale from Subaru overlaid 
on the X-ray image taken by ROSAT All Sky Survey in 0.1-2.4~keV band.
Here, the exposure time was corrected and instrumental backgrounds were subtracted. 
The X-ray image was smoothed by a Gaussian of $\sigma \approx 2 \arcmin$.  
The numbers in color bar is in units of counts~s$^{-1}$~pixel$^{-1}$.
The FOVs of the {\it Suzaku} pointings are plotted with  boxes (yellow).
The solid (white) circles indicate $r_{500}$ ($\sim 44\arcmin$) and $r_{\rm vir}$ 
$\sim r_{98}$ ($\sim 97\arcmin$), respectively, 
which were derived with weak-lening observations by \citet{Okabe2010}.
The dashed (magenta) box shows the field of the Subaru weak-lening survey \citep{Okabe2013} 
and the mass distribution derived from this survey was overlaid in contours.
The  crosses (green) show the positions of 
representative galaxies, NGC~4807, the NGC 4816 group, and IC~4088, 
for the subhalos ``ID~1", ``ID~2" and ``ID~32", respectively.
(A color version of this figure is available in the online journal.)
}
\label{fig:rosatimage}
\end{center}
\end{figure*}

In the Coma cluster, \citet{Okabe2013} detected three massive subhalos
whose mass are higher than $10^{13}~M_{\sun}~h^{-1}$, and which are labeled
as ``ID~1",  ``ID~9", and ``ID~32".
The weak-lensing signal of these three subhalos were well represented
by a truncated NFW model.
Since the ``ID~9" is associated with a halo of the famous subgroup around NGC~4839, 
which has been already observed with $Suzaku$ and reported by \citet{Akamatsu2013},
we observed ``ID~1",  ``ID~32", and a southern offset region of ``ID~32" as a background 
(hereafter ``ID~32 BGD") with {\it Suzaku}.
We also observed ``ID~2" subhalo 
,which is associated with the NGC~4816 group
and the total mass reach to 9$\times 10^{12}~M_{\sun}~h^{-1}$, with {\it Suzaku}.
The mass, truncation radius (hereafter $r_t$),  
and coordinates of each subhalo are summarized 
in table \ref{tb:subhalodata} and observational logs with {\it Suzaku} 
are shown in table \ref{tb:obslog}.
The three subhalos, "ID~1", "ID~2", and "ID~32", 
are located at the projected distances of 
1.4 $r_{500}$, 1.2 $r_{500}$,  and 1.6 $r_{500}$ 
from the X-ray peak of the Coma cluster, respectively.
In figure \ref{fig:rosatimage}, we overlaid the field of views (FOVs) of
{\it Suzaku} pointings and the contours of mass map derived from 
 weak-lensing \citep{Okabe2013} on the X-ray image with {\it ROSAT}.
We also used  four {\it Suzaku} pointings beyond $2.5~r_{\rm 500}$ of 
the Coma cluster to study the X-ray background emission.
The details are described  in Appendix~\ref{sec:cxb}.

In this study, we used only XIS data.  
The XIS instrument consists of three sets of X-ray CCDs (XIS~0, 1, and 3).  
XIS~1 is a back-illuminated (BI) sensor, while XIS~0 and 3 are front-illuminated (FI).  
The instruments were operated the normal clocking mode (8 s exposure per frame).
The data were reprocessed the standard screening criteria
\footnote{http://heasarc.nasa.gov/docs/Suzaku/processing/criteria\_xis.html}
using HEAsoft 6.15. 
We also performed event screening with the cosmic-ray cut-off rigidity COR $>$ 6 GV, 
and the Earth rim ELEVATION $>$10\degr.
We generated Ancillary Response Files (ARFs) by 
``xissimarfgen'' Ftools task \citep{Ishisaki2007}, assumed a 
uniform sky of 20$\arcmin$ radius.  
The effect of degrading energy resolution by radiation damage 
was included in the redistribution matrix files by ``xisrmfgen'' 
Ftools task.  We employed the night Earth database generated by the 
``xisnxbgen'' Ftools task for the same detector area to subtract 
the non-X-ray background (NXB).

\section{ANALYSIS AND RESULTS}
\subsection{X-ray images and Surface brightness profiles} 
\label{sec:surface}

\begin{figure*}[htpd]
\begin{center}
\includegraphics[width=0.5\textwidth,angle=90,clip]{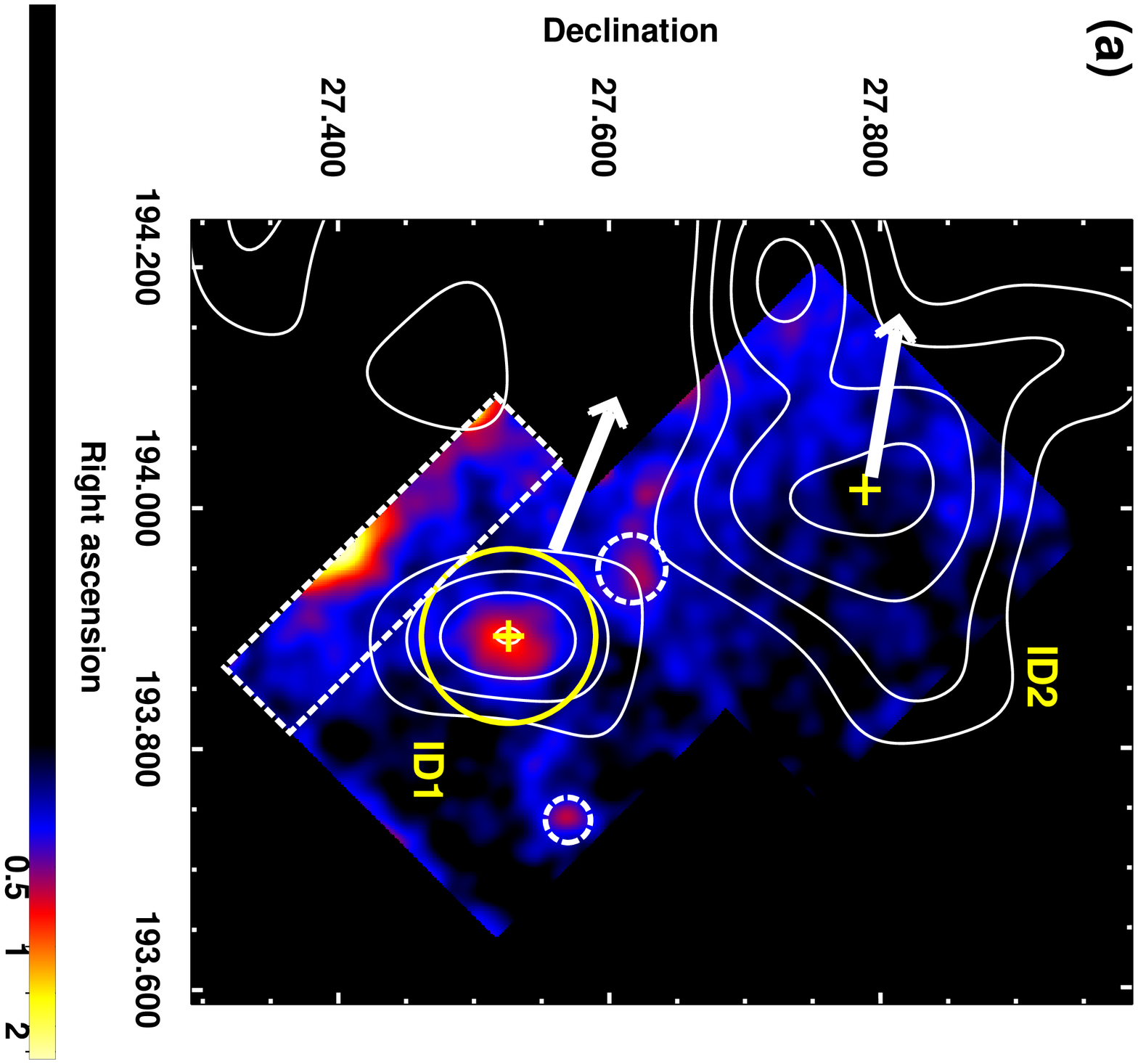}
\includegraphics[width=0.5\textwidth,angle=90,clip]{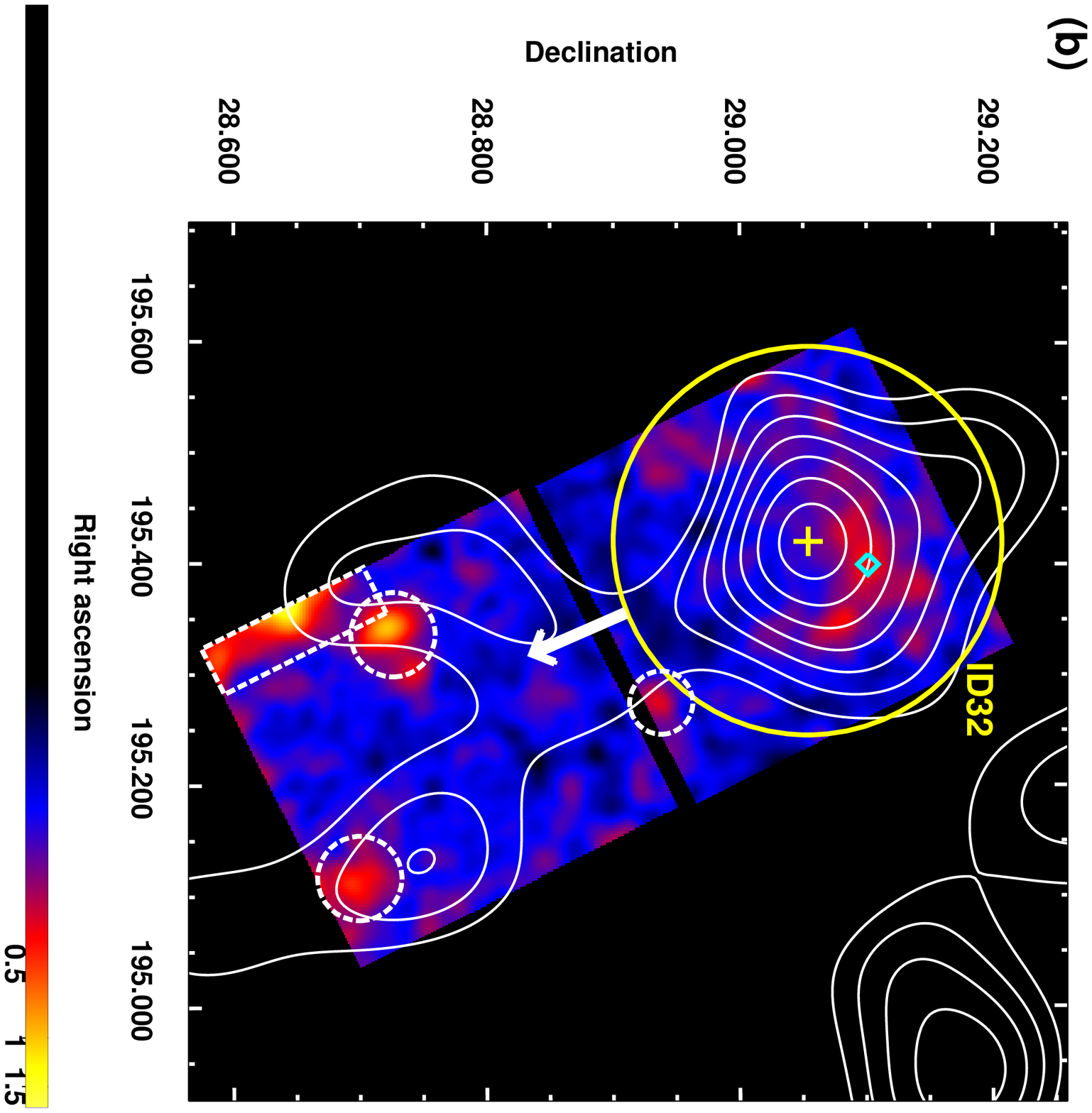}
\caption{
The weak-lensing mass contour map in linear scale overlaid 
on the NXB subtracted XIS images in 0.5--2.0 keV.
(a) around ``ID~1" and ``ID~2", and (b) around ``ID~32" in the 0.5--2.0 keV energy band\@.
Here, the difference in the exposure times and the vignetting effect at 1~keV were corrected.
The images were smoothed by a Gaussian of $\sigma=$24 pixels $\approx 25\arcsec$.  
The numbers below the color bars have units of counts Ms$^{-1}$ pixel$^{-1}$.
The contours (white) show the mass map derived from weak-lening by \citet{Okabe2013}.
The direction of the Coma cluster center from the center of 
each subhalo is shown by the arrow.
The crosses are the center of each subhalo as summarized in table \ref{tb:subhalodata}.
For "ID~1" and "ID~32", 
the solid (yellow) circles show the truncation radii.
The diamond (cyan) mark in the
 "ID~32" subhalo corresponds to the X-ray peak of the excess emission. 
When creating the surface brightness profiles,
the regions in the dashed (white) circles and boxes were excluded.
(A color version of this figure is available in the online journal.)
}
\label{fig:image}
\end{center}
\end{figure*}

\begin{figure*}[htpd]
\begin{center}
\includegraphics[width=0.4\textwidth,angle=0,clip]{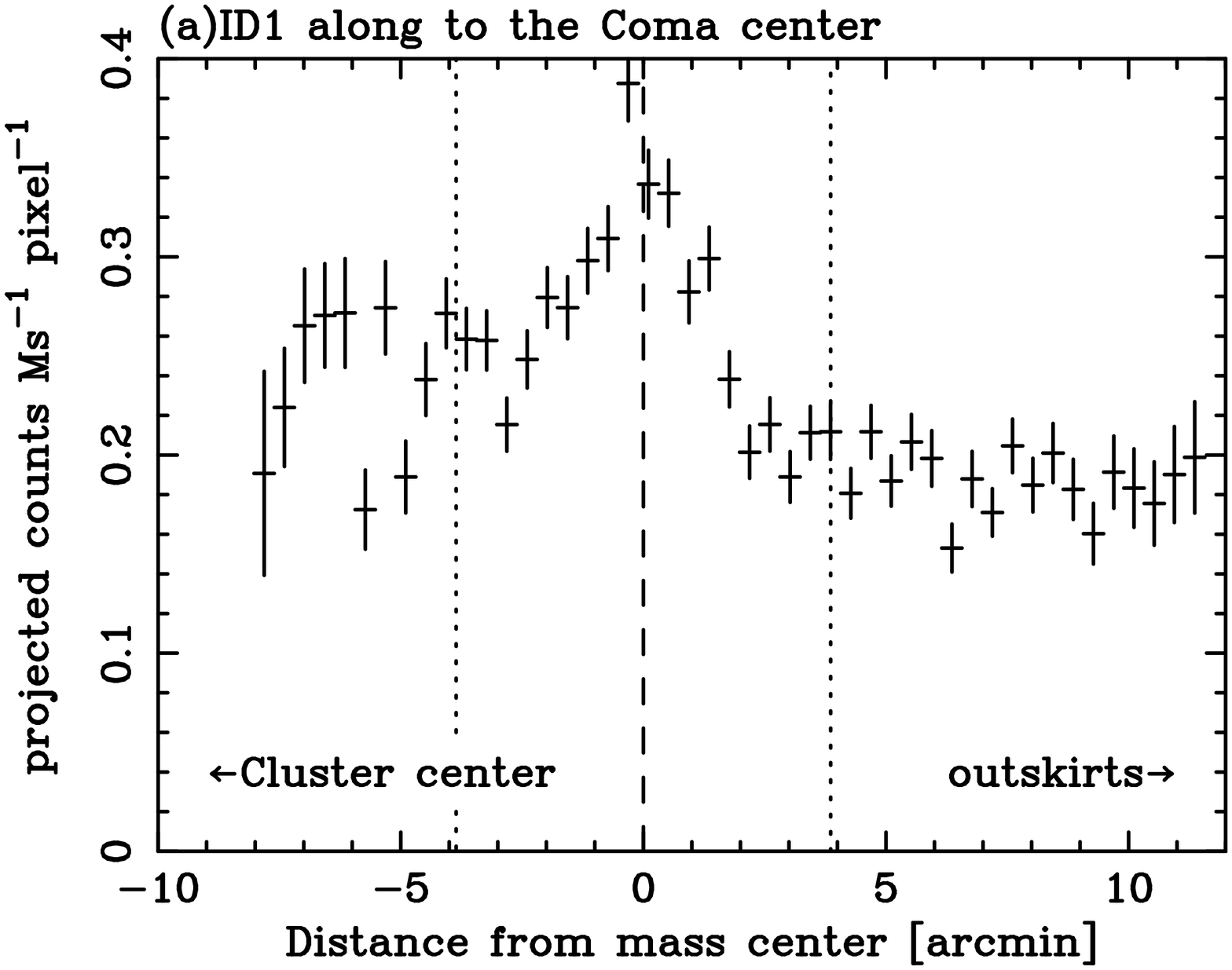}
\includegraphics[width=0.4\textwidth,angle=0,clip]{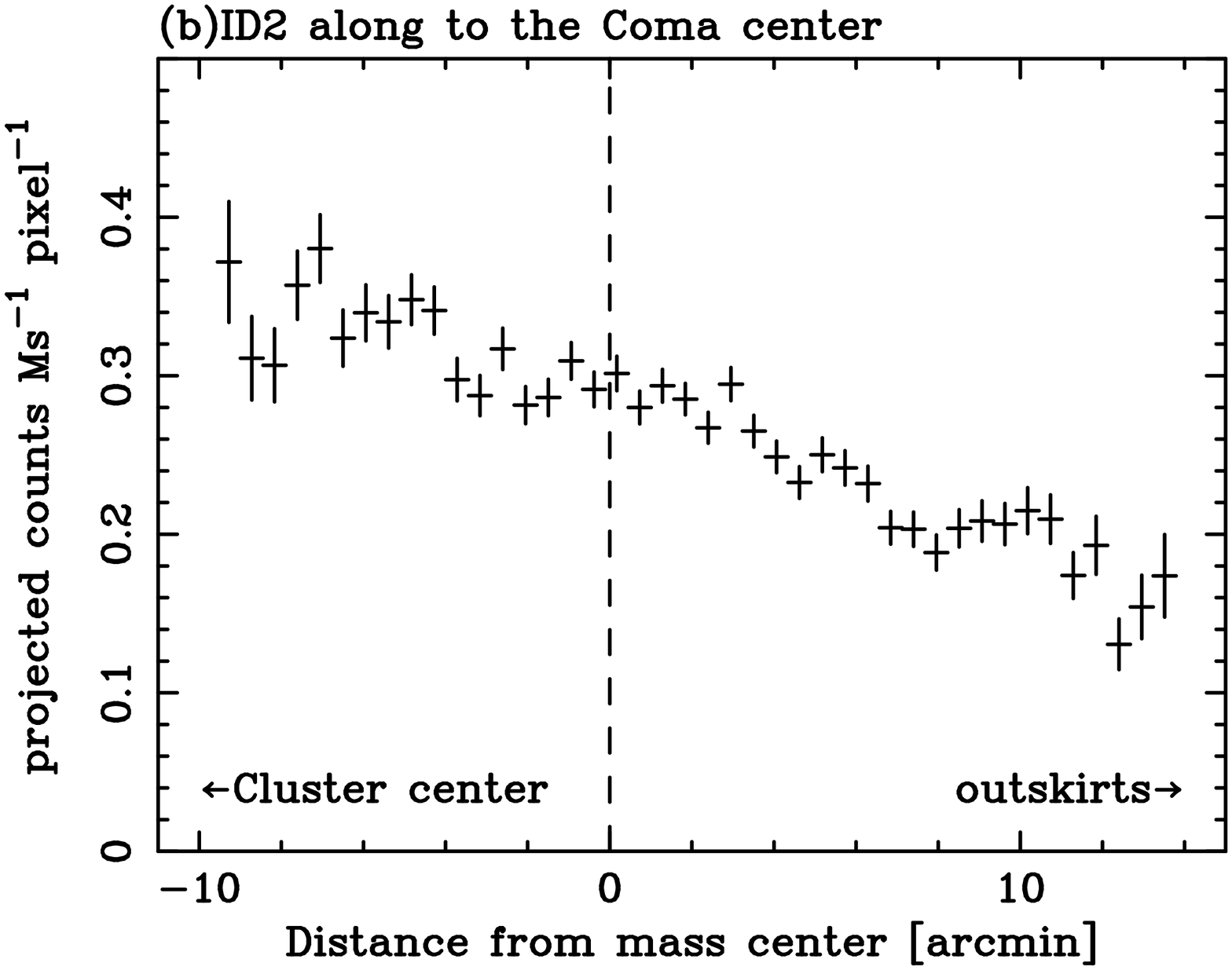}
\includegraphics[width=0.4\textwidth,angle=0,clip]{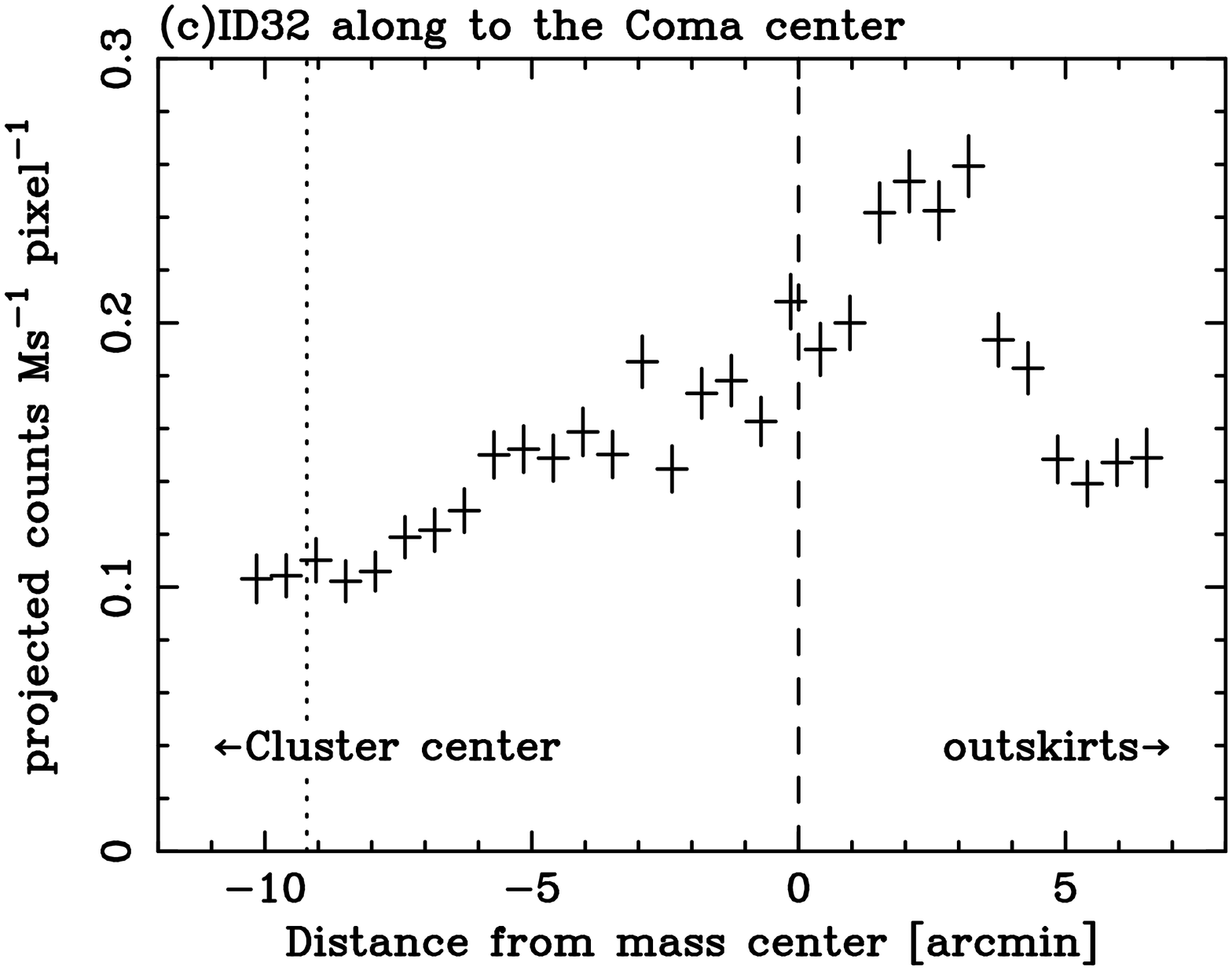}
\includegraphics[width=0.4\textwidth,angle=0,clip]{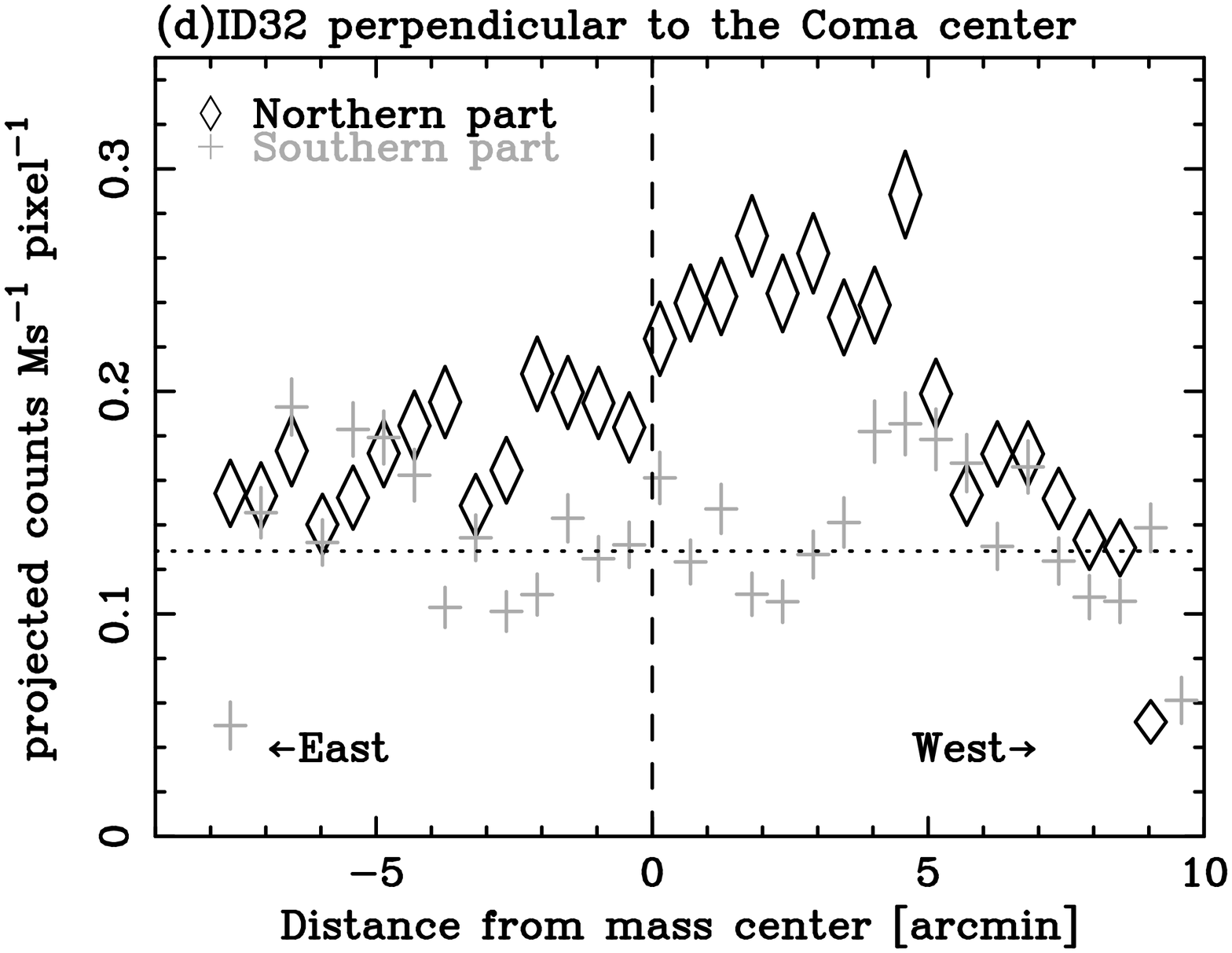}
\caption{
The projected surface brightness profiles of 
(a) "ID~1" subhalo, (b) "ID~2" subhalo, and (c) "ID~32" subhalo
along the direction to the outskirts from the Coma cluster center
extracted from the combined XIS FOV images in the 0.5--2.0 keV energy band.   
The side of negative number corresponds to the direction to the
center of the Coma cluster.
Here, the NXB were subtracted and the effect of the 
the exposure time and vignetting were corrected.
The minus X-axis corresponds to the direction of the Coma cluster center.
The dashed vatical lines indicate the center of each subhalo and
the dotted vatical lines show
the truncation radius of the subhalo ``ID~1" and ``ID~32".
(d)The projection direction is perpendicular against (c) .
The sides of negative and positive number correspond
 to the direction to east and west.
}
\label{fig:SB}
\end{center}
\end{figure*}

In figure \ref{fig:image}, we present combined XIS 
images of subhalos in an 0.5--2.0 keV energy band.
Here, the difference in the exposure times are corrected with exposure map 
generated by ``xisexpmapgen" Ftools task.
In addition, we also corrected the vignetting effect using a flat image at 1~keV\footnote{http://heasarc.gsfc.nasa.gov/docs/suzaku/analysis/expomap.html}. 
We created surface brightness profiles from these images
of individual subhalos along the direction to the center of the Coma cluster, 
 (R.A., Decl.) = ($12^{\rm h}59^{\rm m}44\fs81, 27{\degr}56{\arcmin}49\farcs92$).
The resultant surface brightness profiles are shown in figure \ref{fig:SB}.

As shown in figure \ref{fig:image} (a) and figure \ref{fig:SB} (a), an excess emission is seen 
around the center of the "ID~1" mass contour.
The surface brightness profile shows that most of the excess emission is confined within 
$\sim$ 2\arcmin ($0.6~r_t$) from the mass center.
In contrast, the "ID~2" subhalo
 does not show any excess emission:
the brightness profile gradually increases toward the Coma cluster center
as shown in figure \ref{fig:SB} (b).


In figure \ref{fig:SB} (c), the "ID~32" subhalo shows an excess emission 
whose peak is not located on the center of the mass contour, 
but shifts about 3\arcmin~($0.3~r_t$) away from the subhalo center  
toward the northern part or the opposite direction from the Coma cluster center.
The excess emission is extended to at least 5\arcmin ($0.6~r_t$) from the mass center.
Because of the asymmetrical profile for the ``ID~32", 
we also made  projections of the surface brightness of the
northern and southern parts from the mass center of the subhalo
 along the perpendicular direction against in figure \ref{fig:SB} (c).
 The resultant projections of the surface brightness are shown in figure \ref{fig:SB} (d).
The excess emission at the northern direction extends out to 5--6\arcmin ~($0.6~r_t$)
toward the east and west directions.
The peak of the X-ray emission is located in the northern part 
and shifts about 3\arcmin~($0.3~r_t$) away from the subhalo center
toward west direction.
In contrast, the brightness of most of  the  southern part is consistent with
that of the  background region, although 
there are two peaks at $\sim6\arcmin$ offsets of east and west directions.
We extracted spectra around these two excess. Since their spectra are relatively hard
and fitted with a power-law model, the southern peaks are possibly caused by
background point sources. 

\subsection{Spectral fitting}
\label{sec:results}

\begin{figure*}[t]
\begin{center}
\includegraphics[width=0.45\textwidth,angle=90,clip]{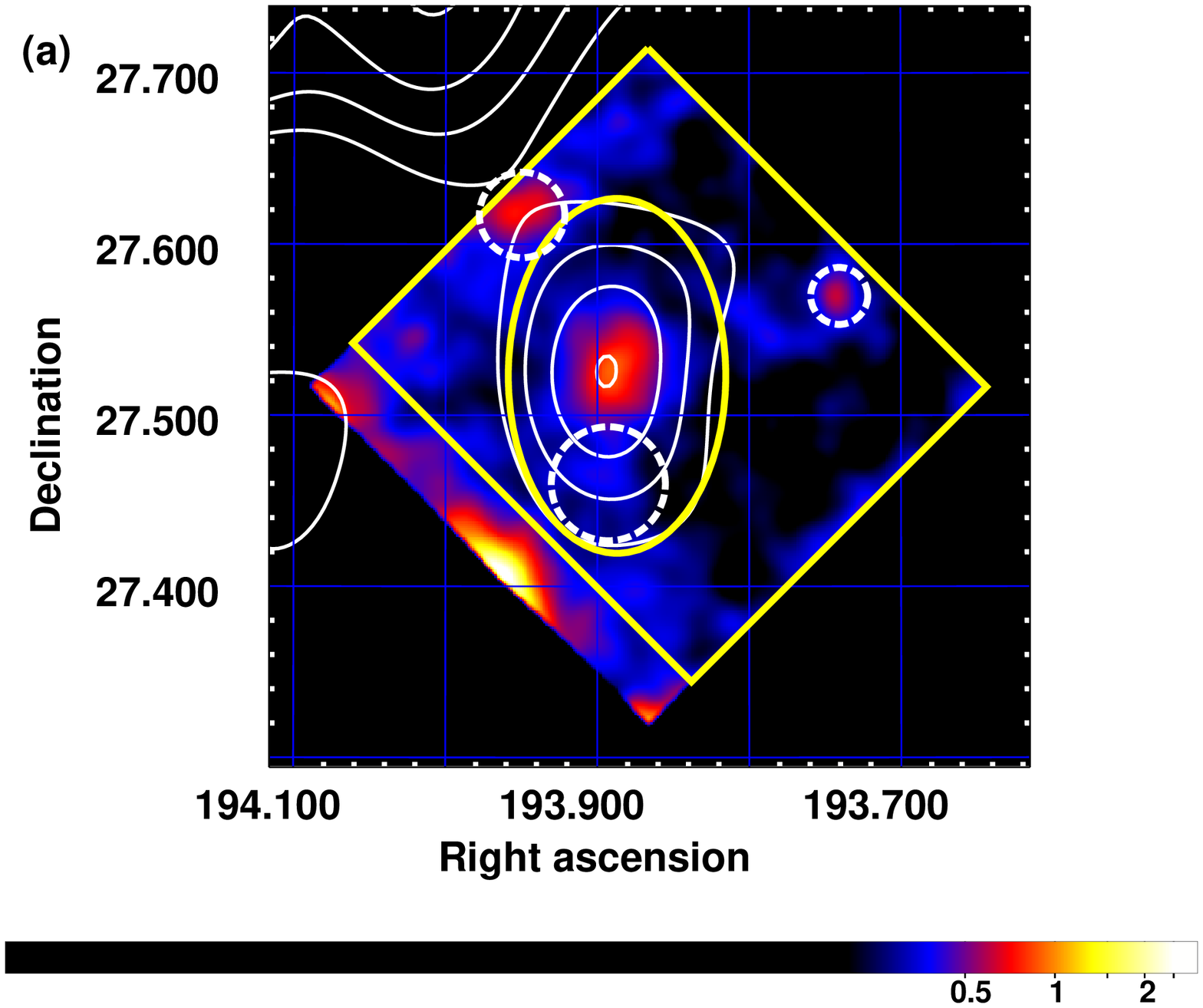}
\includegraphics[width=0.45\textwidth,angle=90,clip]{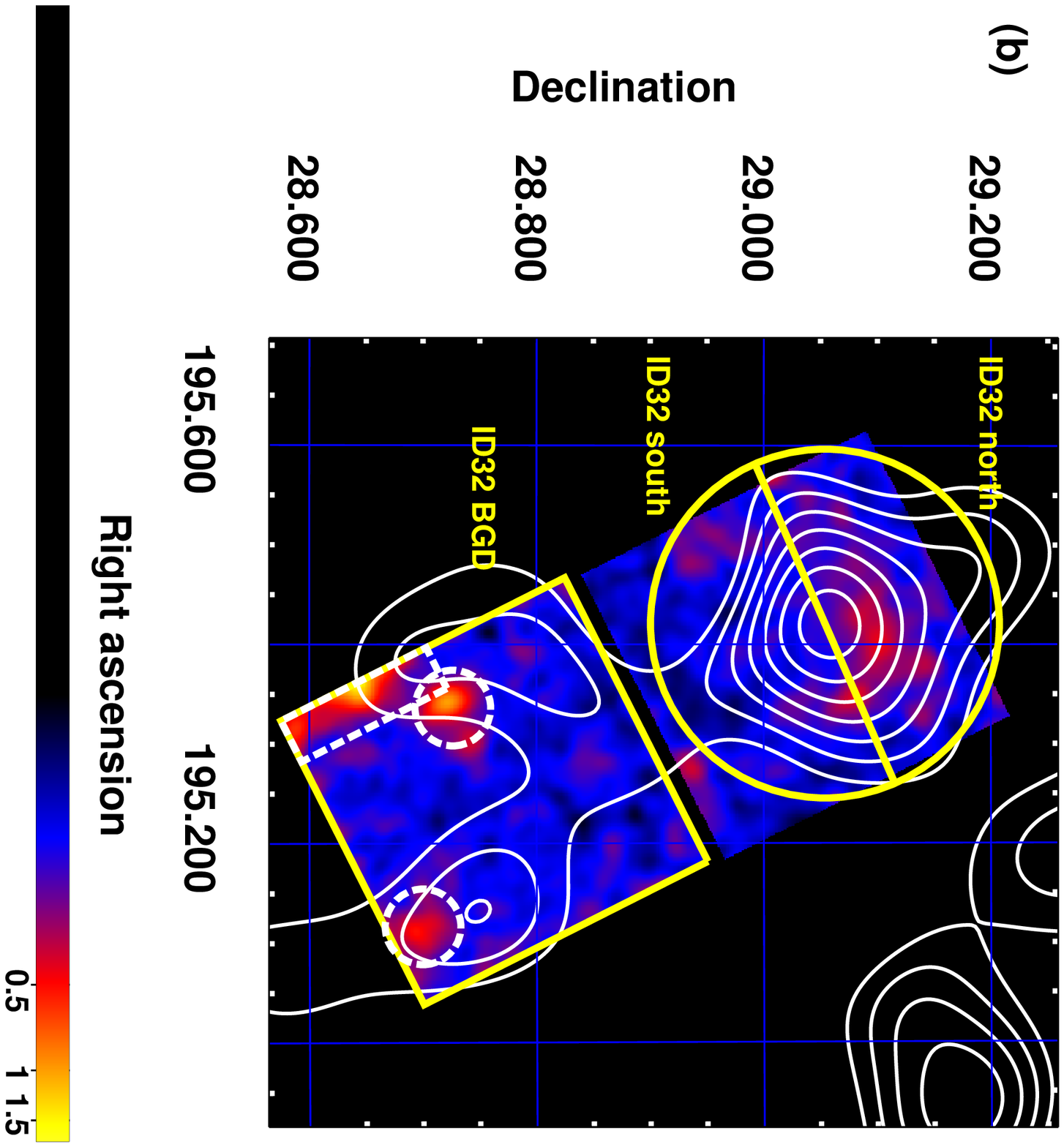}
\caption{The same images as  figure \ref{fig:image}.
The elliptical region (yellow) in the left panel  and semicircular regions (yellow) in the right panel 
are used in spectral extraction for the "ID~1" subhalo, 'south` and 'north`
regions of the "ID~32" subhalo, respectively.
The solid box (yellow) in the left panel excluding the elliptical region and that in the right panel 
are used in the background analysis ("ID~1 BGD" and "ID~32 BGD", respectively).
The regions within dashed box and circles (white) are excluded from these spectral analysis.
(A color version of this figure is available in the online journal.)
}
\label{fig:imageforspec}
\end{center}
\end{figure*}

We extracted spectra over subhalo regions and background regions
as shown in figure \ref{fig:imageforspec}.
The regions around point sources brighter than 
$1\times10^{13}~{\rm erg~s^{-1}~cm^{-2}}$ in 2.0-10.0~keV
were excluded from the spectral analysis.
Since the mass contours of the subhalo ``ID~1" in linear scale are elongated, 
we extracted spectra over an elliptical region,
whose semiminor and semimajor axes are 1.0 and 1.6~${r_t}$, respectively.
Here, we excluded  a circular region around a background galaxy group \citep{Okabe2013},
plotted as a dashed circle in figure \ref{fig:imageforspec}.
To study background emissions including the ICM contribution,  
we extracted spectra over an square region (hereafter "ID~1 BGD"), 
excluding the elliptical region.
For the ``ID~32" subhalo, we extracted  spectra over two semicircular regions 
(hereafter 'south` or 'north` regions)  of the subhalo out to the truncation radius.
The background spectra for the ``ID~32" were extracted from the 
FOV of the "ID~32 BGD" observation.

The spectral fitting was carried out using XSPEC 12.8.1g 
and the extended C-statistic estimator.
The spectra were binned to have at least one count per channel.
We used the energy ranges of 0.5--7.0 keV and 
0.7--7.0 keV for the BI and FI detectors, respectively.  
We excluded the energy band around the Si-K edge (1.82--1.84 keV) 
because its response was not modeled correctly.  


We assumed that the X-ray emissions from the "ID~1 BGD"
and "ID~32 BGD" consist of 
the Galactic emissions from the Local Hot bubble (LHB) and the Milky Way Halo (MWH), 
the Cosmic X-ray background (CXB), and the ICM of the Coma cluster. 
The LHB and MWH were modeled with a thermal plasma model
({\it apec} model; \citealt{Smith2001}) without and with 
the Galactic absorption, $apec_{\rm LHB}$ and $phabs_{\rm GAL} \times apec_{\rm MWH}$,
respectively.
Here, the {$phabs_{\rm GAL}$} indicates the Galactic absorption using
{\it phabs} model in the XSPEC package,
and the column density of each subhalo direction 
were summarized in table \ref{tb:subhalodata}. 
The temperature of the $apec_{\rm LHB}$ was fixed at 0.1~keV, 
while its normalization was a free parameter. 
We also allowed the temperature and normalization of 
the $apec_{\rm MWH}$ to vary. For the LHB and MWH, 
the redshift and abundance were fixed to be 0 and 1~solar, respectively.
The CXB emission was described an absorbed power-law model 
with a photon of index $\Gamma = 1.4$, $phabs_{\rm GAL} \times power-law_{\rm CXB}$. 
The normalization of $power-law_{\rm CXB}$ were estimated beyond 
the virial radius of the Coma cluster 
as described in appendix \ref{sec:cxb}.

The ICM emission was modeled with a thermal plasma model
with the Galactic absorption, $phabs_{\rm GAL} \times apec_{\rm ICM}$.
The temperature, normalization, and abundance were allowed to vary,
except for the abundance for "ID~32 BGD" which was fixed at 0.2 solar.
The redshift of the ICM component was fixed at the value of the Coma cluster, $z = 0.0231$.
Thus, we used the following model formula for the spectra for the "ID~1 BGD" and "ID~32 BGD"; 
$constant \times \left( apec_{\rm LHB} + phabs_{\rm GAL} \times 
 \left( apec_{\rm MWH} + apec_{\rm ICM}  
 + power-law_{\rm CXB} \right) \right).$
Here, the $constant$ is a normalization parameter for the difference in relative 
normalizations among XIS detectors.

For the subhalo regions, we added an extra thermal 
plasma model, $phabs_{\rm GAL} \times apec_{\rm subhalo}$.
We allowed the temperature and normalization of 
$apec_{subhalo}$ of the "ID~1" and "ID~32" to vary.
The abundance for the "ID~1`` was a free parameter but that for
the "ID~32" were fixed at 0.2 solar, 
since we can not constrain the abundance. 
Even if the abundance was fixed to be 0.1 or 0.3 solar, the results did not change.  
The redshift of the $apec_{subhalo}$ was
also fixed at the value of  the Coma cluster.
Thus, we modeled the spectra extracted from subhalo regions as following formula; 
$constant \times \left( apec_{\rm LHB} + phabs_{\rm GAL} \times 
 \left( apec_{\rm MWH} + apec_{\rm ICM}  + apec_{\rm subhalo} 
 + power-law_{\rm CXB} \right) \right).$
We finally fitted the spectra extracted for each subhalo region 
and corresponding background region simultaneously, assuming that 
the X-ray background components have 
the same surface brightness, temperature and abundances.
Here, relative normalizations of three XIS detectors were allowed to vary.

\begin{figure*}[htpd]
\begin{center}
\includegraphics[width=0.4\textwidth,angle=0,clip]
{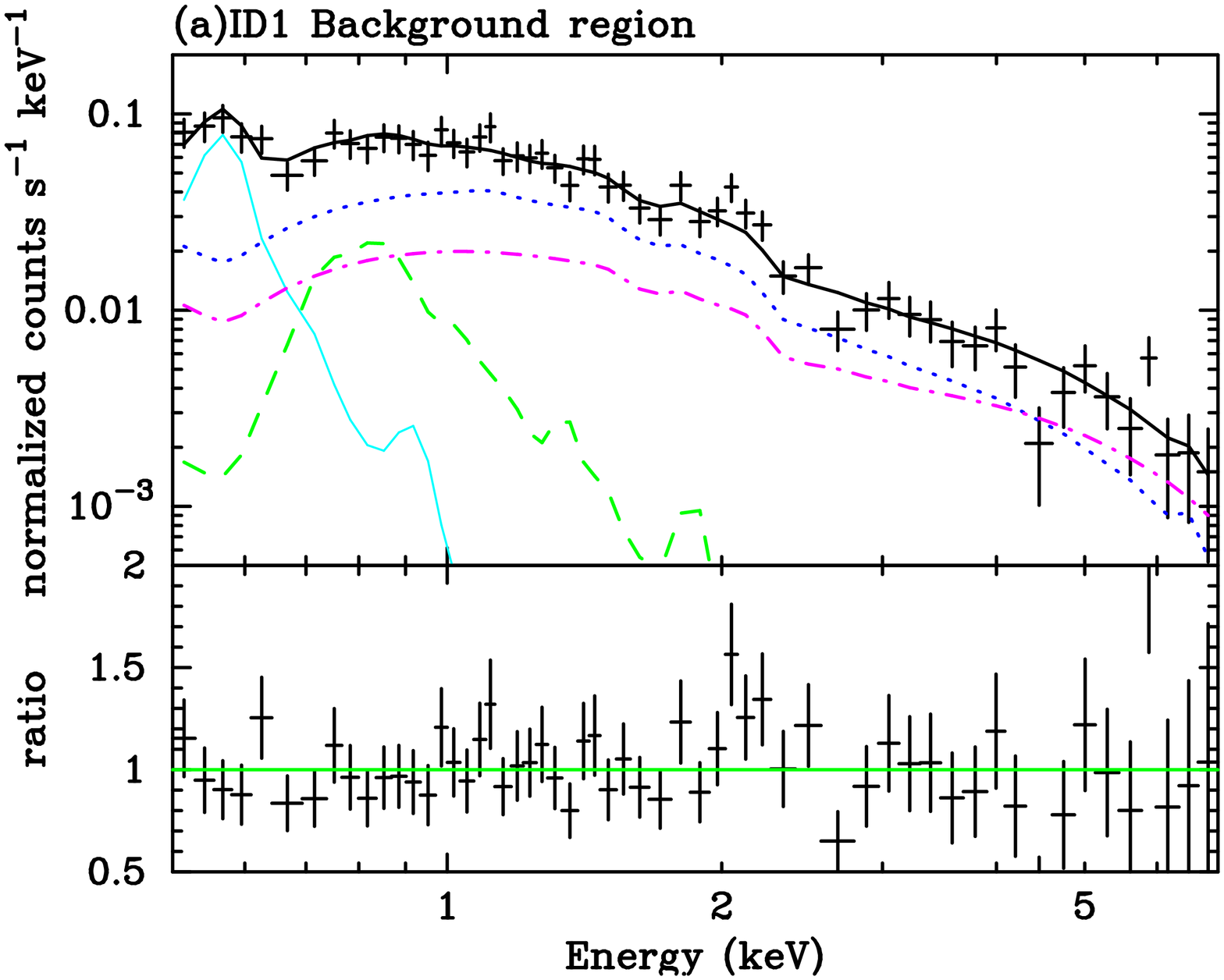}
\includegraphics[width=0.4\textwidth,angle=0,clip]
{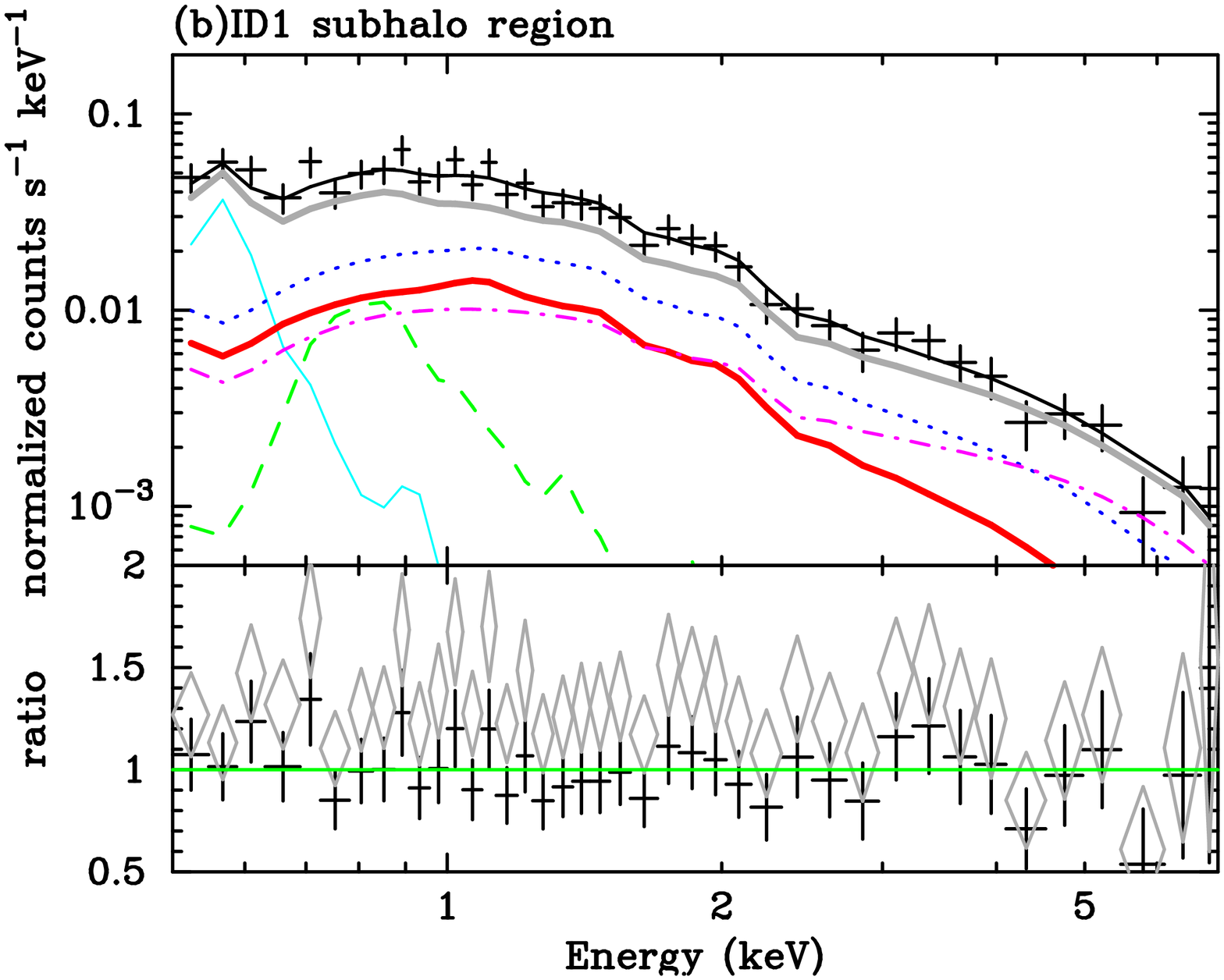}
\includegraphics[width=0.40\textwidth,angle=0,clip]
{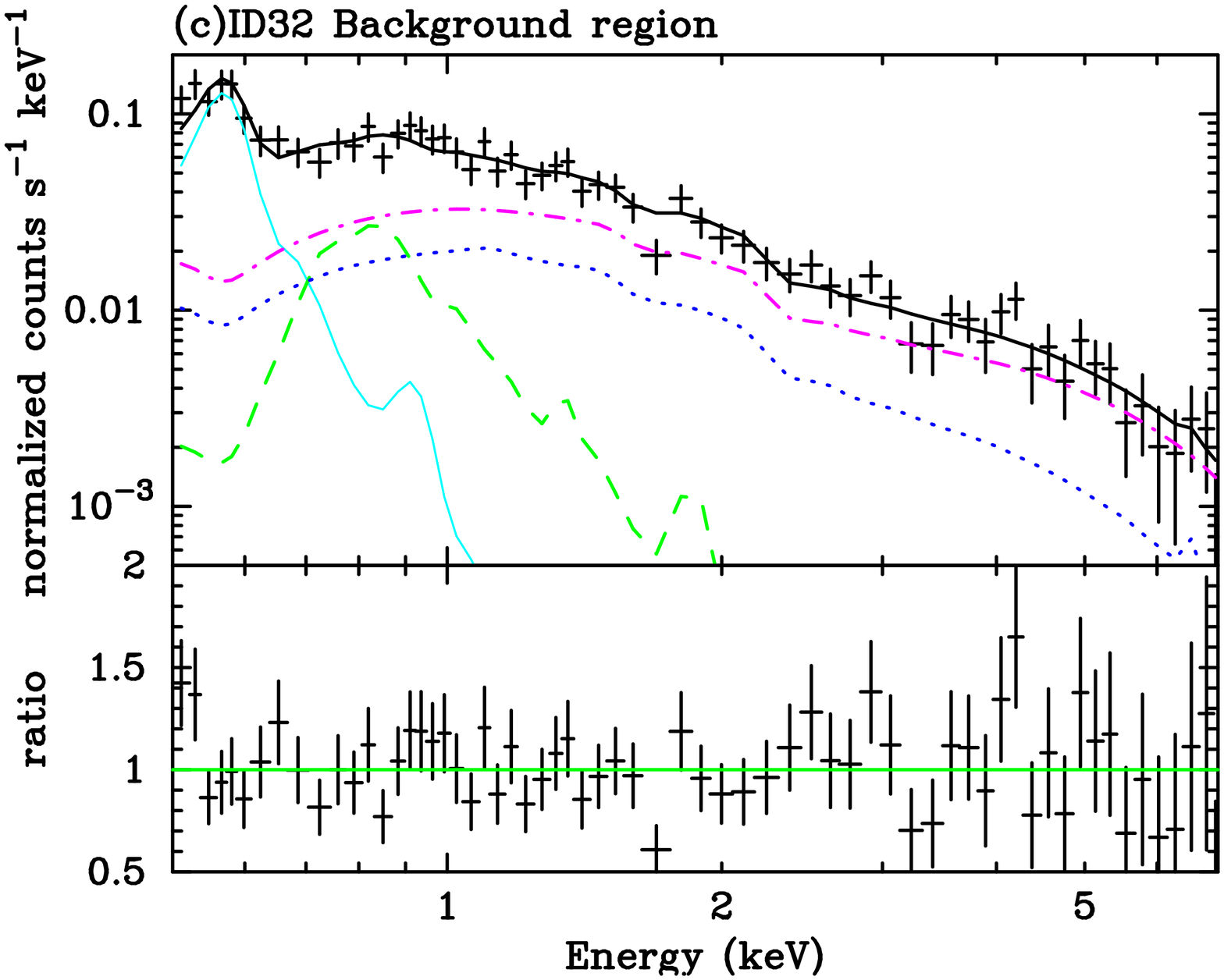}
\includegraphics[width=0.40\textwidth,angle=0,clip]
{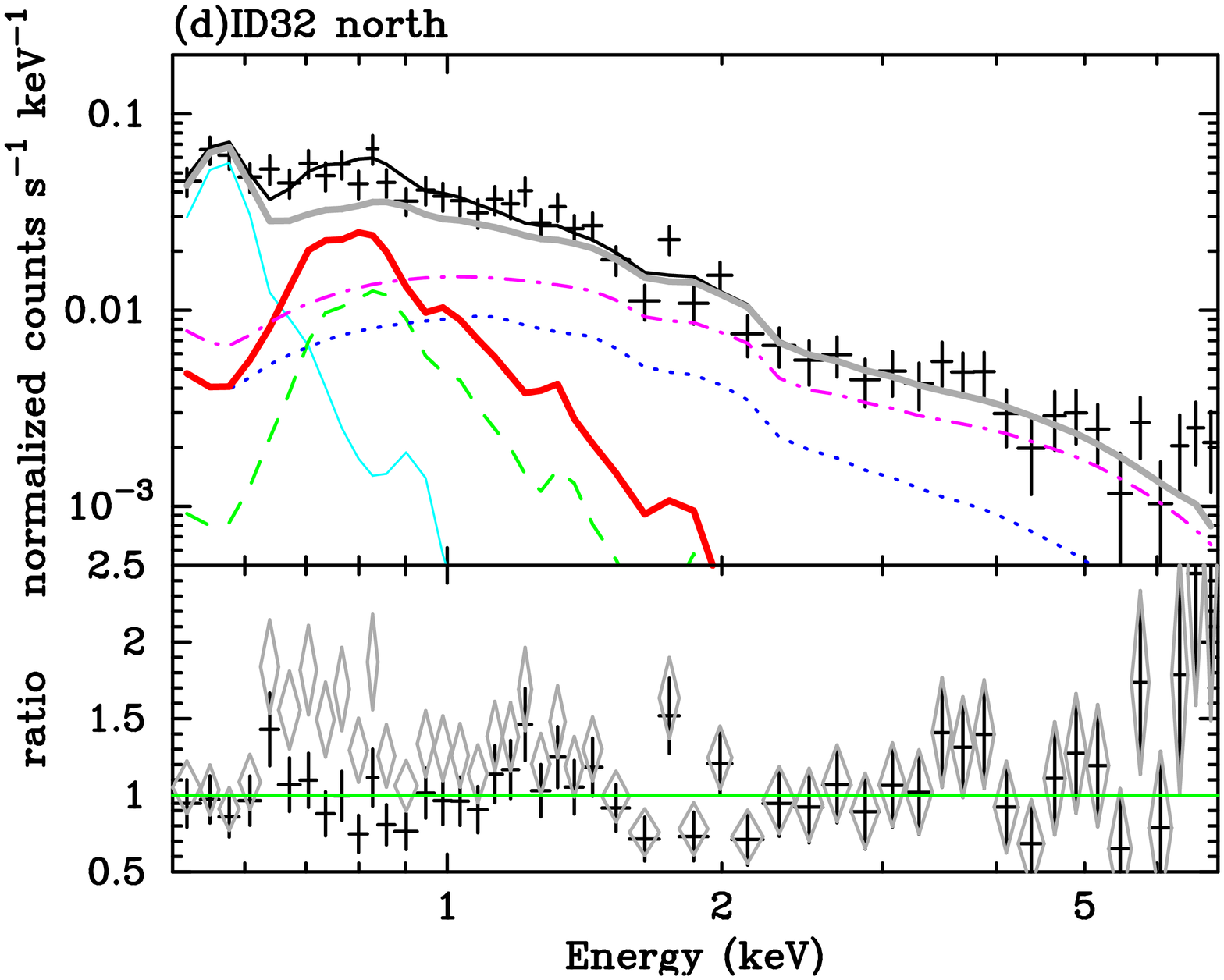}
\includegraphics[width=0.40\textwidth,angle=0,clip]
{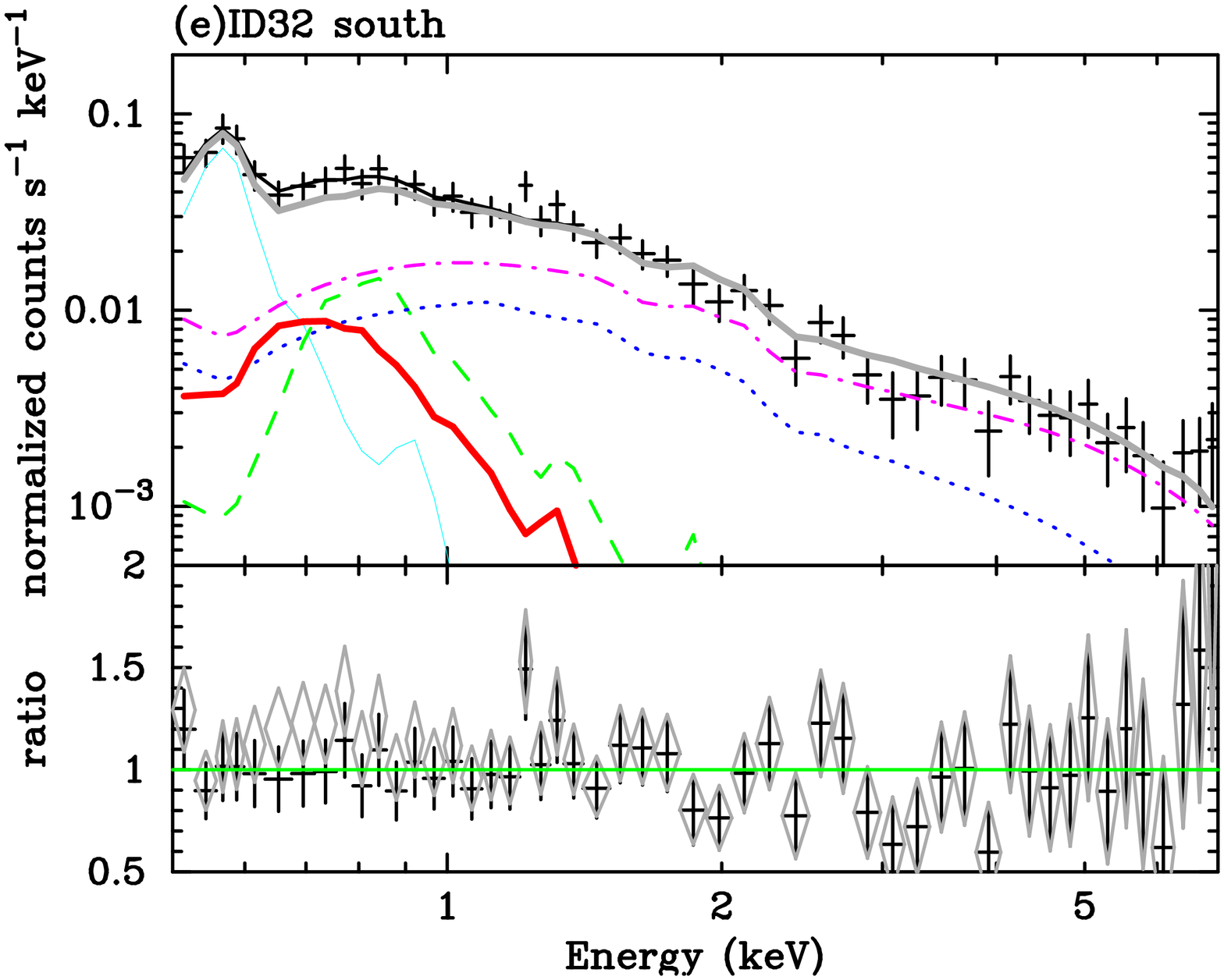}
\caption{
The spectra of (a) ``ID~1 BGD", (b) the subhalo region spectra of ``ID~1",
(c) ``ID~32 BGD", (d) and (e) 'North' and 'south' regions of the subhalo ``ID~32", respectively. 
The spectra were rebinned here for display purpose only.
Upper panels show the NXB subtracted XIS~1 spectra (black crosses).  
The subhalo component is plotted by (red) bold line.
The ICM, CXB, LHB, and MWH components are indicated by 
(blue) dotted, (magenta) dash-dotted, (cyan) thin, and (green) dashed lines, respectively,
and the gray solid line indicates the sum of these background components.
 The lower panels show the data-to-model ratios.
 The gray diamonds in the lower panels in (b), (d) and (e) show those without
 the subhalo component.
In the lower panels in (b), (d) and (e), the gray diamonds show that 
data-to-model ratio without subhalo component, the ratio indicates that 
the subhalo component can not be negligible.  
(A color version of this figure is available in the online journal.)
}
\label{fig:spec}
\end{center}
\end{figure*}

The fitting results are summarized in table \ref{tb:results}.
Figure \ref{fig:spec} shows the best fit spectra of the background regions. 
The ICM temperatures for the ``ID~1" and ``ID~32" regions,
  $4.33^{+0.60}_{-0.37}$~keV and $5.19^{+1.04}_{-0.83}$ keV, respectively,
are consistent with previous results of southwest and northwest 
directions \citep{Simionescu2013} at similar distance from the cluster center, respectively.
Although the error range is fairly large, the ICM abundance of the ``ID~1"
was also consistent with that in \citet{Simionescu2013}.

The temperatures and normalizations of the Galactic 
components are consistent between the ``ID~1" and ``ID~32".
The normalization of the LHB derived from the "ID~1" and "ID~32" spectral fits
is consistent with that for the region  at $110\arcmin-130\arcmin$ (Appendix).
However, the temperature and normalization of MWH are significantly different
from those derived for the $110\arcmin-130\arcmin$  region.
When we use the temperature and normalizations of the 
the Galactic components obtained from $110\arcmin-130\arcmin$ region, 
and fitted the spectra of the subhalo regions, 
the temperature and normalization of the "ID~1" and the 'north' region 
of the "ID~32" did not change within statistical errors, 
although the temperature of the  'south' region of "ID~32" decreased to about 0.1~keV.
Considering the possible spatial variation of the Galactic components,
we adopted the results of the simultaneous fits using "ID~1~BGD" and "ID~32~BGD".


\begin{deluxetable}{lllllll}
\tabletypesize{\scriptsize}
\tablewidth{0pt}
\tablecaption{
The fitting spectral results. 
\label{tb:results}
}
\tablehead{
 \multicolumn{7}{c}{Background components} \\ \tableline
\colhead{Field name} & \colhead{$Norm_{LHB}^{a}$} & \colhead{$kT_{MWH}$}  & \colhead{$Norm_{MWH}^{a}$} & \colhead{$kT_{ICM}$}  & \colhead{$Z_{ICM}$}  &  \colhead{$Norm_{ICM}^{a}$} \\
\colhead{} & \colhead{} &  \colhead{keV} & \colhead{}  & \colhead{keV} & \colhead{Solar} & \colhead{} } 
\startdata
ID 1 BGD  & 18.1$^{+2.0}_{-1.9}$& $0.60^{+0.11}_{-0.09}$ & $0.41^{+0.08}_{-0.07}$  & 4.33$^{+0.60}_{-0.37}$ & 0.11$^{+0.17}_{-0.11}$ & 8.87$^{+0.57}_{-0.65}$  \\
ID 32 BGD&  18.1$^{+1.2}_{-1.2}$& $0.61^{+0.06}_{-0.05}$ & $0.30^{+0.05}_{-0.05}$  & $5.19^{+1.04}_{-0.83}$ & 0.2(fixed)  & 2.68$^{+0.20}_{-0.19}$  \\   \tableline \tableline
 \multicolumn{7}{c}{Subhalo components} \\ \tableline
 & redshift  & $kT$  & $Z$  &  $Norm^{a}$  & Luminosity$^b$ & Flux$^b$   \\
 & & keV  & solar  &    & 10$^{41}~{\rm erg~s^{-1}}$ &   10$^{-13}~{\rm erg~s^{-1}~cm^{-2}}$\\ \tableline
%
``ID~1"& 0.0231 &$2.71^{+0.99}_{-0.59}$ & $0.13^{+0.36}_{-0.13}$ & $5.72^{+1.32}_{-1.32}$ & $1.78^{+0.39}_{-0.33}$ &  $1.48^{+0.24}_{-0.29}$\\
``ID~32" north & 0.0231& $0.55^{+0.07}_{-0.13}$ & 0.2(fixed) & $2.80^{+1.00}_{-0.42}$ & $1.76^{+0.17}_{-0.40}$ & $1.44^{+0.14}_{-0.32}$ \\
``ID~32" south & 0.0231& $0.29^{+0.13}_{-0.07}$ & 0.2(fixed) & $2.05^{+1.58}_{-1.06}$ & $0.67^{+0.29}_{-0.48}$ & $0.54^{+0.21}_{-0.41}$ \\ \tableline
%
``ID~1" & 0.418& $3.83^{+1.02}_{-0.82}$ & $0.09^{+0.24}_{-0.09}$ & $9.83^{+1.82}_{-1.71}$ & $(8.38^{+1.39}_{-1.03}) \times 10^{2}$ & $1.50^{+0.21}_{-0.15}$ \\
``ID~32" north & 0.189& $0.90^{+0.07}_{-0.13}$ & 0.2(fixed) & $3.51^{+0.51}_{-0.51}$ & $(1.74^{+0.28}_{-0.30}) \times 10^2$ &  $1.66^{+0.35}_{-0.20}$\\
``ID~32" south & 0.189& $0.68^{+0.19}_{-0.41}$ & 0.2(fixed) & $1.09^{+3.65}_{-0.41}$ & $(5.24^{+1.85}_{-3.52}) \times 10^1$ & $0.49^{+0.28}_{-0.32}$\\ 
\enddata
\tablenotetext{a}{
The normalization of the apec components divided the solid angle, $\Omega^{U}$, assuming a uniform sky of 20$\arcmin$ radius, 
$Norm = \int n_{\rm e} n_{\rm H} dV \,/~\,[4\pi\,(1+z)^2 D_{\rm A}^{~2}] \,/\, \Omega^{U}$ $\times 10^{-17}$ cm$^{-5}$~400$\pi$~arcmin$^{-2}$.} 
\tablenotetext{b}{The energy range is 0.5--2.0 keV.}
\end{deluxetable}

\subsection{Fitting results of the subhalo components}

Figure \ref{fig:spec} also shows the best fit spectra of the subhalo regions.
The fitting results of the subhalo components are summarized in table \ref{tb:results}.
The spectra for the  subhalo and background regions were well-represented 
with our model formula.
Without the subhalo components, the data-to-model ratios show significant excess.

The temperatures of each subhalo component are  lower than that of the  surrounding  ICM.
For the ``ID~1", the temperature is $2.71^{+0.99}_{-0.59}$ keV and is
 cooler than the ICM component, $4.33^{+0.60}_{-0.37}$ keV.
The temperatures of `north' and `south' of  the ``ID~32" are about 
$0.55^{+0.07}_{-0.13}$~keV and $0.29^{+0.13}_{-0.07}$~keV, respectively.
These values are significantly lower than the  surrounding ICM temperature, $5.19^{+1.04}_{-0.83}$~keV.
The abundance of the subhalo ``ID~1" component is consistent with that of the surrounding ICM.
The luminosity of the ``ID~1" component is about $2 \times 10^{41}~\rm erg~s^{-1}$ 
 at 0.5--2.0 keV energy range and those of the 
'north` and 'south` regions  of ``ID~32" are $\sim 2 \times 10^{41}~\rm erg~s^{-1}$ 
and $\sim 7 \times 10^{40}~\rm erg~s^{-1}$  at the same energy range, respectively.

%
\subsection{The representative background structure fitting}

Background galaxy groups are located within the ``ID~1" and ``ID~32" subhalo regions
at $z=0.418$ \citep{Wen2009} and $z=0.189$ \citep{Hao2010}, respectively
\citep{Okabe2013}.
If the weak-lensing signals were mostly caused by the corresponding background galaxy groups,
the virial mass, $M_{\rm vir}$  would be 1--3$\times 10^{15}~ M_\odot$ for 
"ID~1" and several times $10^{14}~ M_\odot$ for "ID~32" \citep{Okabe2013}.
For such massive clusters, we expect that the  ICM temperatures and ICM luminosities
exceed 10 keV and $10^{45}~{\rm erg~s^{-1}}$ for "ID~1", respectively,
and several keV and several times $10^{44}~{\rm erg~s^{-1}}$ for "ID~32", respectively.
Therefore, we refitted the $Suzaku$ spectra  using redshifts of the 
background galaxy groups.
As a result,  the temperature  increased to $\sim 4$~keV and 0.8 keV for
"ID~1" and the 'north' region of "ID~32", respectively,
and the X-ray luminosity became $\sim 10^{44}~ \rm erg~s^{-1}$ 
and  2$\times 10^{43}~{\rm erg~s^{-1}}$ in the 0.5--2.0 keV range, respectively.
These temperatures and X-ray luminosities are far below the expected values
assuming that the weak-lensing signals come from the background galaxy groups.
Thus, it is unlikely that the excess emissions and weak-lensing signals
 come from the background
galaxy groups, and fairly are associated with the Coma cluster.

\subsection{The gas mass estimation}
\label{sec:gasmass}

To estimate the gas mass of each subhalo, 
we first approximated spherical symmetry 
and assumed constant density up to  2.5 arcmin 
and 6 arcmin for ``ID~1" and ``ID~32", respectively
, which correspond to $\sim 0.6~r_{t}$ for each subhalos, since
beyond these radii, the excess emission was not detected.
The best-fit normalization in the spectral fitting leads us to derived 
average electron density within the extracted region of each subhalo. 
The resultant average electron densities of  ``ID~1", 'south` and 'north` regions
of  ``ID~32" are $(5.3\pm0.6)\times10^{-4}$ cm$^{-3}$, $(2.0\pm0.8)\times10^{-4}$ cm$^{-3}$, 
and $(2.4\pm0.4)\times10^{-4}$ cm$^{-3}$, respectively.
Integrating the electron densities out to 0.6~$r_t$, 
the derived gas mass are $(2.1\pm0.2)\times10^{10}~M_{\sun}$,
$(6.8\pm1.2)\times10^{10}~M_{\sun} $, and 
$(5.8\pm2.2)\times10^{10}~M_{\sun}$ for "ID~1" and  the 
'north` and 'south` regions of ``ID~32", respectively.

We also estimated the electron density profiles 
by deprojecting radial profiles of the surface brightness centered on 
the center of the mass contour and the X-ray peak of 
"ID~1" and "ID~32" subhalos, respectively.
We integrated the electron density profiles out to $0.6~r_t$, within which 
the X-ray emissions of each subhalo are detected.
Then, the derived gas mass of "ID~1" and "ID~32" are $(1.5\pm0.1)\times10^{10} M_{\sun}$
and $(9.5\pm0.5)\times10^{10} M_{\sun}$, respectively.
Comparing these values with those assuming constant electron density,
the systematic uncertainties in the gas mass would be about a factor of 2--3.

\begin{figure*}[htpd]
\begin{center}
\includegraphics[width=0.45\textwidth,angle=0,clip]{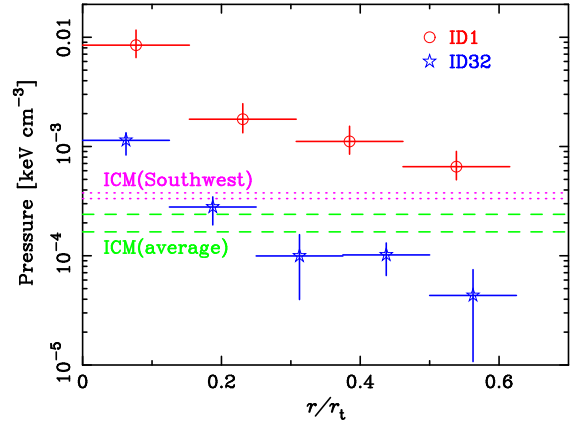}
\caption{
The thermal pressure profile of "ID~1" centered on the mass center (open circles)
and that of "ID~32" centered on the X-ray peak (open stars).
The  dotted (magenta) and dashed (green) lines indicate the ICM pressure range
of the southwest direction and azimuthal average excluding the southwest direction
derived by  \citet{Simionescu2013}.
(A color version of this figure is available in the online journal.)
}
\label{fig:pressure}
\end{center}
\end{figure*}

Using the radial profiles of the electron density profile and the 
temperatures of "ID~1" and the 'north` region of
"ID~32" derived from the spectral fits, 
radial profiles of the thermal gas pressure, $P = n_{e}kT$ out
to $\sim 0.6~r_{t}$, were estimated and plotted 
in figure \ref{fig:pressure}.
Since the surface brightness and temperature of the ICM component surrounding
"ID~1" and "ID~32" are close to those of the southwest direction 
and the azimuthal averages  excluding the southwest
direction derived by  \citet{Simionescu2013}, we compared our pressure profiles
with those derived by  \citet{Simionescu2013}.
The thermal pressure at $0.5~r_t$ of "ID~1" is slightly higher
than that of the southwest direction. 
That in 0.3--0.5 $r_t$ of "ID~32" is slightly below the ICM pressure.


We also estimated the gas mass assuming that excess emission come from 
background galaxy groups located within the ``ID~1" and ``ID~32" subhalo regions.
The calculation methods are the same as mentioned in the previous paragraphs. 
Assuming a flat density profiles, the total gas mass for ``ID~1" and ``ID~32" are 
$(1.8\pm0.2)\times10^{13}~M_{\sun}$ and $(1.3\pm0.8)\times10^{13}~M_{\sun}$, respectively.
By deprojecting the surface profiles, we estimated the gas mass by integrating the calculated density profiles.
The resultant gas masses are $(1.3\pm0.1)\times10^{13}~M_{\sun}$ and $(1.3\pm0.1)\times10^{13}~M_{\sun}$
for ``ID~1" and ``ID~32", respectively.

\section{DISCUSSION}
\label{sec:discussion}

We observed three massive subhalos detected by the weak-lening
survey with Subaru, ``ID~1", ``ID~2" and ``ID~32", 
which are located on the projected distances of 
1.4~$r_{500}$,  1.2~$r_{500}$, and  1.6~$r_{500}$ 
from the center of the Coma cluster, respectively, with {\it Suzaku}.
The excess X-ray emission have been detected from ``ID~1" and ``ID~32", while ``ID~2"
 has no significant excess emission. 
Temperatures of these subhalos  were lower  than that of the surrounding ICM. 
In section \ref{sec:comparison}, we compare the $M_{\rm gas}$-$M_{\rm total}$ and 
$kT$-$M_{\rm total}$ relations between subhalos and regular galaxy groups.
We estimate the ram pressure and mass loss rate caused by the Kelvin-Helmholtz instability in section \ref{sec:stripping}. 
As discussed in \cite{Simionescu2011}, the clumping in the galaxy clusters 
lead us to overestimate the electron density. 
In section \ref{sec:clump}, we study the effect of the subahlos on the density and 
temperature measurements.

\subsection{Comparison of the gas mass fraction and temperature with other clusters}
\label{sec:comparison}

\begin{figure*}[htpd]
\begin{center}
\includegraphics[width=0.36\textwidth,angle=0,clip]{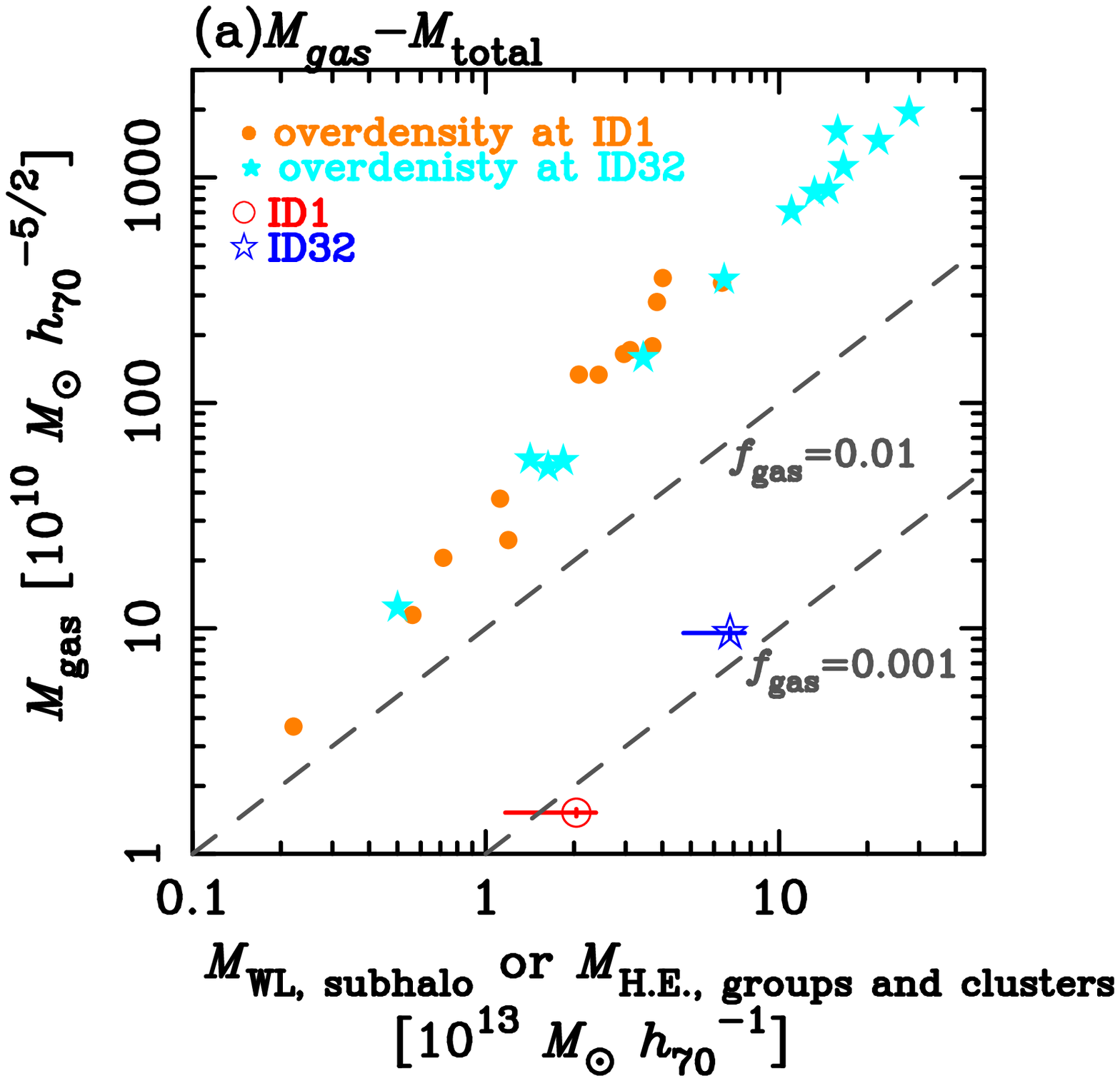}
\includegraphics[width=0.4\textwidth,angle=0,clip]{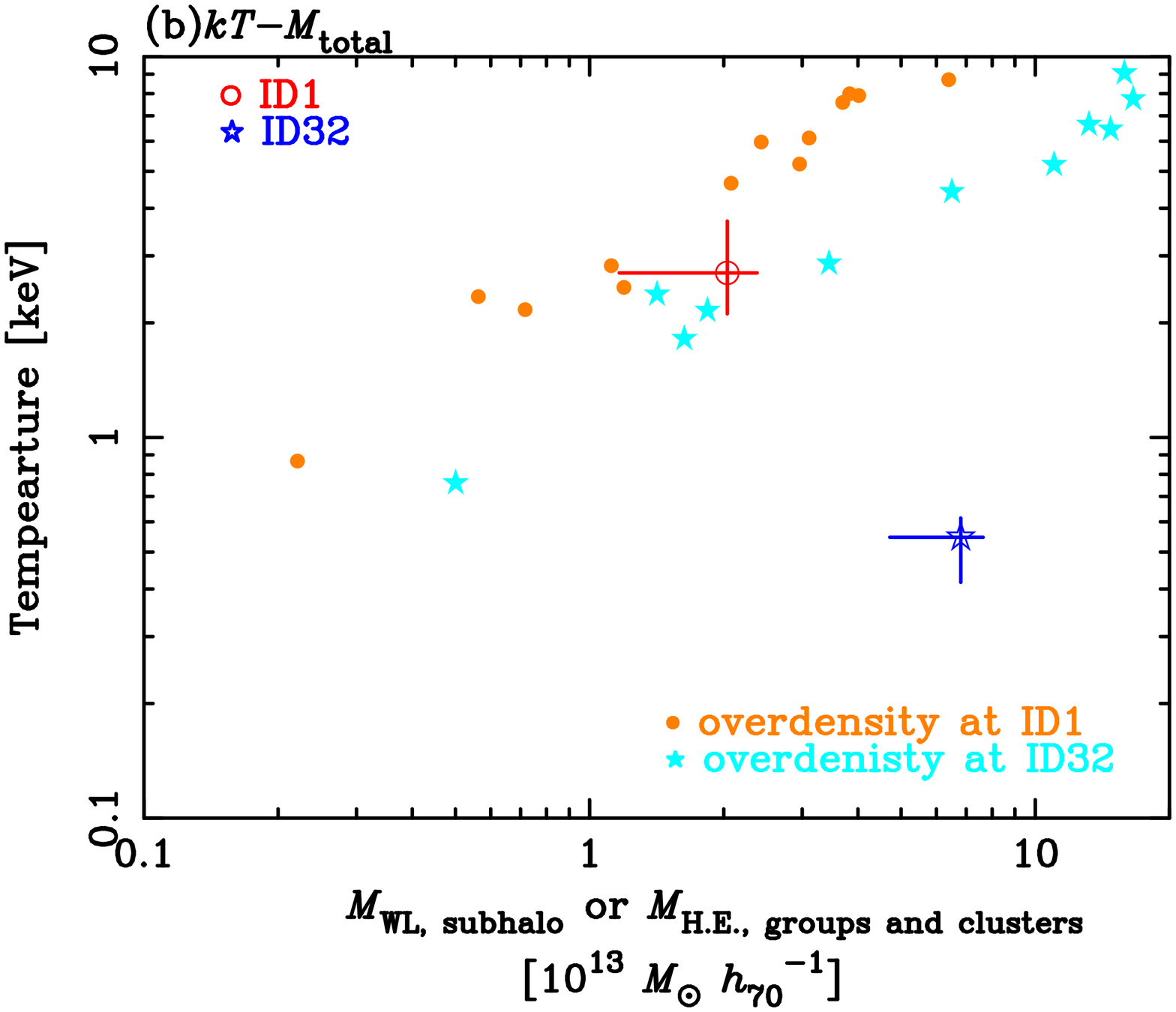}
\caption{
(a) The subhalo gas mass against weak-lening mass or hydrostatic mass. 
The open (red) circle and (blue) star marks indicate ``ID~1" and ``ID~32" , respectively.
The error bars of the  weak-lensing masses of the subhalos include the systematic errors. 
The filled (orange) circles and (cyan) stars are the gas mass vs. hydrostatic mass at 
the radius within which the mean density of the groups and clusters are 
the same overdensity as ``ID~1" and ``ID~32", respectively. 
The gas mass and hydrostatic mass of the groups and clusters 
were calculated from Chandra's results \citep{Vikhlinin2006}.
(b) The same as left panel but temperature vs weak-lensing mass or hydrostatic mass.
}
\label{fig:relation}
\end{center}
\end{figure*}

The gas mass and weak-lensing mass of "ID~1" and "ID~32" are plotted
in figure \ref{fig:relation} (a).
The gas mass fraction,  or the gas mass to weak-lensing mass ratio, of
these two subhalos are about 0.001.
Here, we used the gas mass derived from integrating
 the radial profiles of electron density.
We compared these gas mass fraction of the subhalos
 with the gas mass to hydrostatic mass 
ratios of clusters and groups of galaxies.
Since the  mean density within the truncation radius of subhalo 
``ID~1" and ``ID~32" are about 6600 times and 27000 
times higher than the critical density of the Universe, respectively,
we calculated gas mass and hydrostatic mass at a radius within 
which each average density is the same as each subhalo's overdensity,
using Chandra results by \citet{Vikhlinin2006}.
The gas mass fractions of these clusters correlate well with the hydrostatic mass,
and are about 0.02--0.1, which are about 1--2 orders of magnitudes higher than those of subhalos.
If these subhalos had been regular groups before infalling onto the Coma cluster,
they should have lost most of their gas.



In figure \ref{fig:relation} (b),
we also compared the relation of the temperature and total mass 
($kT$--$M_{\rm total}$ relation) of the subhalos and clusters.
Here, we also calculated the temperature and hydrostatic mass
 of the Chandra clusters in \citet{Vikhlinin2006} 
 at the radius of the same over densities, or
 $r_{6600}$ for "ID~1" and $r_{27000}$ for "ID~32".
The temperature of "ID~1" 
is slightly lower than 
expected by the $kT-M_{\rm total}$ relation of clusters.
For ``ID~32", the temperature of the subhalo is several times 
lower than those of the $kT-M_{\rm total}$ relation of clusters.

\subsection{The effect of ram pressure stripping}
\label{sec:stripping}

\begin{figure*}[htpd]
\begin{center}
\includegraphics[width=0.45\textwidth,angle=0,clip]{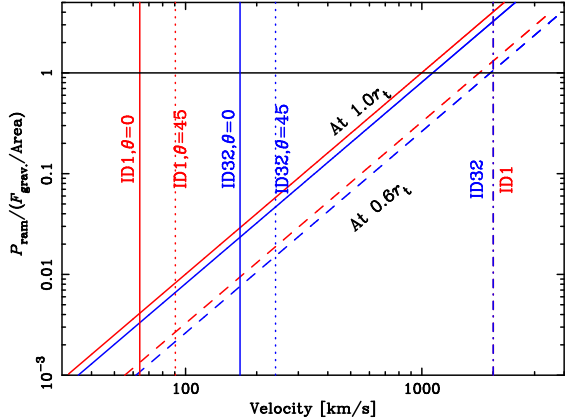}
\caption{
The ratio of the ram pressure to gravitational force per area against the velocity of each subhalo. 
The solid and dashed lines are the ratio calculated at $1.0~r_{t}$ and $0.6~r_{t}$, respectively. 
The red and blue lines correspond to ``ID~1" and ``ID~32", respectively. 
The vertical solid and dotted lines correspond to velocities of each subhalo assuming
 the 0$\degr$ and 45$\degr$ inclination angle, respectively.
The infall velocities are plotted the red and blue dot-dashed lines for ``ID~1" and ``ID~32", respectively.
The horizontal solid line corresponds to unity.
(A color version of this figure is available in the online journal.)
}
\label{fig:ramfraction}
\end{center}
\end{figure*}

The observed very low gas mass fraction and morphologies of the excess
emission indicate that the gas in the infalling subhalos has been stripped via
ram pressure of the surrounding ICM. 
The subhalo would be unable to hold the interstellar materials 
when the ram pressure exceeds the gravitational restoring force per area.
At the truncation radius, the fraction of the ram pressure to the gravitational restoring force per area \citep{Takizawa2006}, 
\begin{equation}
Ratio = \frac{P_{ram}}{F_{grav} / Area}
= 1.3 \left( \frac{r_t}{100~\rm kpc} \right)^4 
\left( \frac{n_{e,\rm ICM}}{ 10^{-5}~\rm cm^{-3}} \right) 
\left( \frac{M_{\rm subhalo}}{10^{13}~M_{\sun}} \right)^{-1}
\left( \frac{M_{\rm gas, subhalo}}{10^{10}~M_{\sun}} \right)^{-1}
\left( \frac{v}{ 2500~\rm km~s^{-1} }\right)^{2}
\end{equation}
is the useful parameter to investigate the 
present stripping effect. 
Here, the $n_{e,ICM}$, $v_{\rm gal}$, $r_t$,  $M_{\rm subhalo}$, and $M_{\rm gas, subhalo}$ 
are the electron density of the ICM, velocity of the subhalo, truncation radius, mass of subhalo, 
and gas mass of subhalo, respectively.
By adopting those parameters of the subhalos and ICM, 
we calculated the ram pressure to the gravitational force 
per area ratio as a function of the subhalo moving velocity and plotted
in figure \ref{fig:ramfraction}.
We also estimated the ratio at the $0.6~r_t$, which corresponds to the border of the excess X-ray emission.
The ram pressure should  have been lower than the present value
when infalling the outer regions, since the ICM density decreases with the distance from the cluster center.


Since NGC~4807 and IC~4088 are located near the center of the mass
contours of ``ID~1" and ``ID~32"  (see figure \ref{fig:rosatimage}), respectively, 
 these galaxies are likely representative galaxies of these subhalos. 
The recession velocities of NGC~4087 and IC~4088 are 
$6989\rm~km~s^{-1}$ and $7095\rm~km~s^{-1}$ 
(NASA/IPAC Extragalactic Database\footnote{http://ned.ipac.caltech.edu/}), respectively. 
Considering that of  the Coma cluster, $6925\rm~km~s^{-1}$, 
the velocities in line of the sight of these galaxies are
$64\rm~km~s^{-1}$ and $170\rm~km~s^{-1}$, respectively. 
As shown in figure \ref{fig:ramfraction},
if these subhalos are moving toward the line of sight, 
 the ratio of the ram pressure to the gravitational force 
 in a unit area are orders of magnitudes lower than the unity.
Even if these subhalos are moving with a inclination angle of 45 degree, the ram pressure 
is not still effective to remove the gas of the subhalos.


On the other hand, 
the infall velocities for "ID~1" and "ID~32" are about $2000~\rm km s^{-1}$, which is estimated using the best-fit NFW profile
derived from the weak-lensing for the main halo  \citep{Okabe2010}.
 This value is comparable with the infall velocity of the subcluster, NGC~4839 group, 
whose infall velocity is $1700^{+350}_{-500}$ km s$^{-1}$ \citep{Colless1996}.
Adopting the infall velocity, at $1.0~r_t$, 
the ram pressure is higher than the gravitational force 
per area and enough to remove the gas, and
at  $0.6~r_t$, or at the border of the excess X-ray emission, 
the ram pressure is comparable to the gravitational force per area.
These subhalos are located beyond $r_{500}$ on the sky, and
if they are moving perpendicular to the line of sight with the infall 
velocity, their gas beyond $0.6~r_t$ can be stripped via 
ram pressure stripping.



We also estimated the mass loss rate caused by Kelvin-Helmholtz instabilities 
since the gas in the center of "ID~32" seems to be removed.
\citet{Nulsen1982} estimated the mass loss rate caused 
by viscous stripping via Kelvin-Helmholtz instabilities, 
$\dot{M}_{KH} \approx \pi r_D^2 \rho_{ICM} v$.
Here, $r_D$ and $\rho_{ICM}$ are the disk radius and 
gas density of the ICM, respectively.
The mass loss rate can convert
\begin{equation}
\dot{M}_{KH} \approx 25 \left( \frac{n_e}{\rm 10^{-5}~cm^{-3}} \right) \left( \frac{r_D}{\rm 100~kpc} \right)^2  \left( \frac{v}{\rm 2500~km~s^{-1}} \right) [\rm M_{\sun} yr^{-1}].
\end{equation}
We assumed that $r_{D}$ is same as the truncation radius. 
Using the velocity of line of sight of each subhalo, 
the results of the mass loss rate were the $\dot{M}_{HK} \approx 6~M_{\sun}$ yr$^{-1}$ 
for ``ID~1" and $\approx 60~M_{\sun}$ yr$^{-1}$ for ``ID~32".
In contrast, adopting the infall velocity, the mass-loss rates of ``ID~1" and ``ID~32" are 
180 and 680 $M_{\sun} \rm yr^{-1}$, respectively.
Considering 
 the time scales for the mass-loss of 520~Myr and 390~Myr for "ID~1" and "ID~32", respectively, 
 infalling from the virial radius of the Coma cluster, 
the total  mass loss from adopting the infall velocity
the subhalo are about $9\times10^{10} M_{\sun}$ 
and $3\times10^{11} M_{\sun}$ for ``ID~1" and ``ID~32", respectively.
These values of mass are more massive than current gas mass 
of the both subhalos. 
Therefore,  destroying of gas by Kelvin-Helmholtz instability
 explains the very low gas fraction of "ID~1" and "ID 32" subhalos.
However, the mass-loss rate would be smaller with magnetic fields
which  suppress the destruction by Kelvin-Helmholtz instability.

\subsection{The contribution of the X-ray emission of subhalos to the ICM}
\label{sec:clump}

$Suzaku$ enables us to measure the entropy profiles of the galaxy clusters 
out to the virial radius 
\citep{George2009, Reiprich2009, Bautz2009, Hoshino2010, Kawaharada2010, Simionescu2011, Urban2011, Akamatsu2011, Humphrey2012, Walker2012a, Akamatsu2012, Walker2012b, SatoT2012, Walker2013, Simionescu2013, Ichikawa2013, Urban2014, Sato2014, Okabe2014, Mochizuki2014}.
Contrary to the prediction of 
the accretion shock heating model prediction, the entropy of galaxy clusters 
show flat profiles beyond $r_{500}$.
\citet{Simionescu2011} interpreted that gas density in the outskirts 
are overestimated due to gas clumping and the entropy are underestimated.
However, observing the Perseus cluster outskirts with {\it Chandra},
the number of detected sources is consistent with the background sources
\citep{Urban2014}.
With X-ray and weak-lensing joint analysis, 
\citet{Okabe2014} discussed that entropy flattening of the outskirts of the galaxy clusters 
caused by the steepening of the temperature profiles, rather than the flattening of the gas density.

We studied the effect of subhalo luminosities on the estimation of electron densities
when these subhalos are not excluded from spectral analysis.
The observed X-ray flux of the excess emissions of
 "ID~1" and "ID~32" are (1--2)$\times$ 
$10^{-13}~{\rm erg~s^{-1}{cm}^{-2}}$ in the 0.5--2.0 keV energy band.
Considering that  the threshold of detection of point sources with a 10 ks $Suzaku$ exposure
is about $10^{-13}~{\rm erg~s^{-1}{cm}^{-2}}$ in the 2.0-10.0~keV range,
if similar subhalos are located in other clusters, most of them 
would be below the detection threshold flux of {\it Suzaku}.
We note that regions around subhalos like "ID~9", or the southwest subcluster around 
NGC~4839, can be easily excluded from spectral analysis of clusters observed with {\it Suzaku},
because of their very high X-ray luminosities.

Within projected distance 1.2--1.6 $r_{500}$ where the two subhalos are located,
the ICM gas mass would be several times 10$^{13}~M_\odot$, which is estimated
using the weighted average of the radial profiles of electron density observed
with {\it Suzaku} by \citet{Simionescu2011} excluding
the southwest direction, where the X-ray luminous subhalo, "ID~9" is located.
Thus, the ICM gas mass is two orders of magnitude higher than the gas mass 
of excess emission of the "ID~1" and "ID~32".
The 0.5--2.0 keV luminosities of these two subhalos are a few times 
$10^{41}~{\rm erg~s^{-1}}$ and are negligible when comparing with the X-ray luminosity
of the Coma cluster within 1.2--1.6 $r_{500}$, $\sim 10^{43}~{\rm erg~s^{-1}}$,
which is also estimated using the average of ICM temperature and normalization
observed with {\it Suzaku} \citep{Simionescu2013}
excluding the southwest direction.
Although \citet{Okabe2013} detected 32 subhalos with weak-lensing observations,
most of them are less massive and located within $r_{500}$.
As a result, we can conclude that X-ray emission of subhalos does not affect
the overestimating the gas mass of the Coma cluster.

In order to evaluate the bias in the ICM temperature measurements, 
we extracted spectra again from FOVs observed around each subhalo, 
and fitted the spectra with ICM and background model. 
The temperature of the ICM did not change within statistic error range, since 
the total flux of subhalos is  1-2 orders of magnitude lower than that of the  ICM.
If number of subhalos are much higher, they may affect on the temperature 
and density measurements. Therefore, we created the mock spectra assuming 
two thermal components for clumps and surrounding ICM emission, 
by changing the flux ratio of two components.
%
When we simulated a sum of mock spectra of 2 and 1~keV emissions for 
the ICM and subhalo whose flux is half of that of ICM, respectively,
the derived temperature with a single temperature model became 1.2~keV, 
which is 40\% lower than the original ICM temperature, and the electron density 
was also overestimated by 30\%. 
This indicates the entropy to be low biased by 50\%. 
Thus, if there are many clumps enough to increase the normalization of the ICM 
for changing the entropy profiles, the temperature bias is more significant.

\citet{Simionescu2013} derived the entropy of the Coma cluster beyond $r_{500}$
($\sim 47\arcmin\pm1\arcmin$) derived from \citet{Planck2013}, 
and the entropy profile is consistent with the accretion shock heating model \citep{Pratt2010}.
Here, the $r_{500}$ derived from weak-lening analysis \citep{Okabe2010} is well consistent with 
that from \citet{Planck2013}.
Since there is no evidence for the entropy flattening like other clusters and pressure excess, 
they suggested the gas clump are easily destroyed in a dynamical active cluster.
Therefore, their discussion is consistent with our study that gas clump does not affect
to the gas density estimation of the Coma cluster.

\section{SUMMARY AND CONCLUSIONS}
We observed the three massive subhalos, ``ID~1", ``ID~2", and ``ID~32",  
which are detected from Subaru weak-lening analysis \citep{Okabe2013} with {\it Suzaku}. 
The weak-lensing survey of subhalos in the outskirts of the galaxy cluster enable 
us to efficiently carry out follow-up X-ray observations gas subhalo candidates associated with weak-lensing detected subhalos.

While the excess emission is seen around the center of ``ID~1" mass contour, 
the ``ID~32" subhalo shows that the emission peak is shifted in the northern part or 
outer side from the Coma cluster center.  
In contrast to above two subhalos, the "ID 2" subhalo
does not show any excess emission.
The spectral analysis indicated that the temperature of the subhalo gas is significantly 
lower than the surrounding ICM. 
By deprojecting the surface brightness profiles, we derived 
the gas mass of each subhalo. The total gas mass of ``ID~1" and ``ID~32" are 
$2\times10^{10}~M_{\sun}$ and $1\times10^{11}~M_{\sun}$, respectively.
Comparing with the $kT-M_{total}$ and $M_{\rm gas}-M_{total}$ relation of regular galaxy groups, 
gas fractions of the subhalos are much lower than regular galaxy groups. 
Adopting the infall velocity estimated from the best-fit NFW profile derived by weak-lensing analysis of 
the Coma cluster \citep{Okabe2010}, beyond 0.6 times the truncation radius of the subhalos, or at the border of the excess X-ray emission, the ram pressure is effective to remove the gas.
With the infall velocity, total amount of the removed gas mass from the subhalos via Kelvin-Helmholtz instabilities 
are about $9\times10^{10} M_{\sun}$ and $3\times10^{11} M_{\sun}$ 
for ``ID~1" and ``ID~32", respectively.
The luminosities of subhalos are about two orders of magnitude lower 
than that of the Coma cluster outskirts and do not affect on the gas mass estimate
of the ICM.




\section*{Acknowledgements}
We thank the anonymous referee for careful reading the manuscript and providing valuable comments. 
We also thank all members of the {\it Suzaku} operation team and the XIS calibration team. 
We acknowledge the support of a Grant-in-Aid for Scientific Research from the MEXT, No. 25400235(K. M.), 25800112 (K. S.), and 26800097(N. O.).
This work was supported by ``World Premier International Research Center Initiative (WPI Initiative)" and the Funds for the Development of Human Resources in Science and Technology under MEXT, Japan.

\appendix
\section{The CXB estimation}
\label{sec:cxb}

For estimations of the Cosmic X-ray Background (CXB) level, 
we extracted the spectra beyond $110\arcsec$ offset observations, 
which is corresponding to $2.5~r_{500}$ \citep{Okabe2010}, 
excluding south region from the Coma cluster, 
whose observation logs are shown in table \ref{tb:obslog}.
We searched for point-like sources with ``wavdetect'' tool in 
CIAO\footnote{http://cxc.harvard.edu/ciao/} in 0.5--2.0 and 2.0--5.0 keV images\@.
We also excluded the area around the hot pixels
\footnote{http://www.astro.isas.ac.jp/suzaku/doc/suzakumemo/suzakumemo-2010-01.pdf}.
The flux level of the faintest source was about 1$\times10^{-13}$
erg s$^{-1}$ cm$^{-2}$ in 2.0--10.0 keV with a power-law model 
of a fixed photon index, $\Gamma=$1.7.
We assumed that the background emission were 
composed to two thermal Galactic emissions, 
the Local Hot Bubble (LHB) and the Milky Way Halo (MWH), 
and the Cosmic X-ray background. 
The LHB and MWH were modeled non-absorbed and 
absorbed thermal plasma model ($apec$; \citealt{Smith2001}). 
The CXB was modeled an absorbed power-law model. 
We also convolved the Galactic absorption 
with photoelectric absorption model, {\it phabs}.
The column density was fixed to be $8.5 \times 10^{19}~{\rm cm^{2}}$ 
\citep{Kalberla2005}.
Therefore, we modeled the spectra by the follow formula, 
$constant \times \left( apec_{\rm LHB} + phabs \times 
 \left( apec_{\rm MWH} + powerlaw \right) \right).$
Although the temperature of the LHB 
was fixed at 0.1~keV, the normalization was allowed to vary.
The temperature and normalization of the MWH 
were free parameters. The photon index of the CXB was fixed at 1.4, 
and the normalization was allowed to vary.
The fitting results of are summarized in table \ref{tb:backgrounds}. 
The derived CXB normalization was consistent with \citet{Kushino2002}.
The parameters of  LHB and MWH are comparable with previous results \citep{Simionescu2013}.

\begin{deluxetable}{lllll}
\tabletypesize{\scriptsize}
\tablewidth{0pt}
\tablecaption{
{\it Suzaku} observation logs for the CXB estimation.\label{tb:cxblog}
}
\tablehead{
\colhead{Field name } & \colhead{Sequence No.} & \colhead{Obs. date$^{a}$} &  (R.A., decl.)$^{b}$  &  \colhead{Exposure$^{c}$}  \\
\colhead{} & \colhead{}  & \colhead{} & J2000.0 & ksec  }
\startdata
East 110\arcmin & 806037010 & 2011-06-19T14:09:55 & $13^{\rm h}07^{\rm m}48\fs6$, $27\degr53\arcmin08\farcs5$ & 11.1\\
East 120\arcmin & 806038010 & 2011-06-20T01:23:51 & $13^{\rm h}08^{\rm m}27\fs4$, $27\degr53\arcmin06\farcs0$& 13.8\\
NW  110\arcmin & 806045010 & 2011-06-22T17:48:52 & $12^{\rm h}56^{\rm m}18\fs6$, $29\degr33\arcmin08\farcs3$ & 14.1\\
NW  120\arcmin & 806046010 & 2011-06-23T06:03:44 &$12^{\rm h}56^{\rm m}00\fs8$, $29\degr41\arcmin27\farcs2$ & 11.9\\ 
\enddata
\tablenotetext{a}{Start date of observation, written in the DATE-OBS keyword of the event FITS files.}
\tablenotetext{b}{Average pointing direction of the XIS, written in the RA\_NOM and DEC\_NOM keywords of the event FITS files.}
\tablenotetext{c}{Exposure time after screening.}
\end{deluxetable}

\begin{deluxetable}{lllll}
\tabletypesize{\scriptsize}
\tablewidth{0pt}
\tablecaption{
The background fitting results. 
\label{tb:backgrounds}
}
\tablehead{
 \colhead{$Norm_{\rm LHB}^{a}$} & \colhead{$kT_{\rm MWH}$}  & \colhead{$Norm_{\rm MWH}^{a}$} & \colhead{$Norm_{\rm CXB}^{b}$} \\
 \colhead{} &  \colhead{keV} & \colhead{} & \colhead{} } 
\startdata
$11.4_{-1.4}^{+1.4}$ & $0.31_{-0.02}^{+0.03}$ & $ 0.84_{-0.15}^{+0.15}$ & $1.01^{+0.02}_{-0.02}$  \\  
\enddata
\tablenotetext{a}{The normalization of the apec components divided the solid angle, $\Omega^{U}$, assuming a uniform sky of 20$\arcmin$ radius, 
$Norm = \int n_{\rm e} n_{\rm H} dV \,/~\,[4\pi\,(1+z)^2 D_{\rm A}^{~2}] \,/\, \Omega^{U}$ $\times 10^{-17}$ cm$^{-5}$~400$\pi$~arcmin$^{-2}$.} 
\tablenotetext{b}{The normalization of powerlaw is units of photons cm$^{-2}$ s$^{-1}$ keV$^{-1}$ 400$\pi$ $\rm arcmin^{-2}$ at 1 keV.}
\end{deluxetable}




\end{document}